\documentclass[12pt]{iopart}
\pdfoutput=1
\usepackage{color}
\usepackage{amssymb}

\usepackage{graphicx} 
\usepackage{relsize}

\def\beqra{\begin{eqnarray}}
\def\eeqra{\end{eqnarray}}
\def\beq{\begin{equation}}
\def\eeq{\end{equation}}
\def\etap{\eta^\prime}
\def\etain{\eta_{in}}
\def\ds{\displaystyle}

\def\bk{{\bf k}}
\def\vp{\varphi}

\def\bx{{\bf{x}}}
\def\bp{{\bf{p}}}
\def\bq{{\bf{q}}}
\def\re#1{(\ref{#1})}

\def\half{\mbox{\small$\frac{1}{2}$}}

\def\agt{\stackrel{>}{\sim}}
\def\alt{\stackrel{<}{\sim}}
\begin{document}

\title{Nonlinear Power Spectrum from Resummed Perturbation Theory: a Leap Beyond the BAO Scale}
\author{Stefano Anselmi}
\address{Institut de Ci\`encies de l'Espai, IEEC-CSIC, Campus UAB,\\
Facultat de Ci\`encies, Torre C5 par-2, Barcelona 08193, Spain}
\ead{anselmi@ieec.uab.es}
\author{Massimo Pietroni}
\address{INFN, Sezione di Padova, via Marzolo 8, I-35131, Padova, Italy}
\ead{massimo.pietroni@pd.infn.it}

\begin{abstract}
A new computational scheme for the nonlinear cosmological matter power spectrum (PS) is presented. Our method is based on evolution equations in time, which can be cast in a form extremely convenient for fast numerical evaluations. A nonlinear PS is obtained in a time comparable to that needed for a simple 1-loop computation, and the numerical implementation is very simple. Our results agree with N-body simulations at the percent level in the BAO range of scales, and at the few-percent level up to $k\simeq1\,\mathrm{h/Mpc}$ at $z\agt 0.5$, thereby opening the possibility of applying this tool to scales interesting for weak lensing.
We clarify the approximations inherent to this approach as well as its relations to previous ones, such as the Time Renormalization Group, and the multi-point propagator expansion. We discuss possible lines of improvements of the method and its intrinsic limitations by multi streaming at small scales and low redshifts.
\end{abstract}

\maketitle

\section{Introduction}
Cosmological perturbation theory (PT) as a tool to study the Large Scale Structure of the Universe (LSS)  (for a review see \cite{PT}) has received considerable interest in the recent past. The main motivation is the study of Baryonic Acoustic Oscillations (BAO) imprinted in the matter power spectrum (PS), which are one of the main observables of present and future galaxy surveys \cite{Eis05,2007ApJ...657...51P,GaztaIII,Percival:2009xn,Blake:2011wn,Nuza:2012mw,Sanchez:2012sg,Blake:2012pj,2009arXiv0912.0914L}. The goal of these measurements is to derive the acoustic scale at the percent level accuracy, in order to provide constraints on the Dark Energy equation of state competitive with those obtained from measurements of the magnitude-redshift relation of type Ia Supernovae \cite{Perlmutter98,Riess98}. Of course, on the theory side, the matter PS in the BAO range of scales ($\simeq 100$ Mpc/h) must be computed at the same level. For a $\Lambda$CDM cosmology, the goal can be accomplished by means of high accuracy N-body simulations, once the various issues related to precise initial conditions, very large simulation volumes, mass resolution and time stepping are carefully addressed, as discussed, {\em e.g.}, in \cite{Heitmann:2008eq}. However, the long computational times required by these simulations make it impossible to implement grid based or Markov Chain Monte Carlo (MCMC) estimations of cosmological parameters, which typically require the evaluation of thousands of PS's. Moreover, if departures from $\Lambda$CDM cosmologies are taken into account, for instance by the inclusion of massive neutrinos, non gaussian initial conditions, or $f(R)$ theories of gravity, the N-body approach is still far from being firmly established at the percent level in the relevant range of scales. 

The usage of PT can potentially help in most of these respects. First, being based on analytic, or semi-analytic, techniques, computational times are in general greatly reduced with respect to N-body simulations. Moreover, although the basic formalism is derived for an Einstein-de Sitter cosmology, its extension to $\Lambda$CDM is straightforward, and also its formulations in non-standard cosmologies are feasible and under control \cite{Pietroni08, Koyama:2009me, Scoccimarro:2009eu, Sefusatti:2011cm, Anselmi:2011ef, SefuDami}.  However, the real boost to these methods was given by the papers of Crocce and Scoccimarro (CS) \cite{RPTa,RPTb}, who showed how the poorly behaving PT series can be reorganized, in what they named ``renormalized PT" (RPT), in such a way as to obtain a better behaved expansion, valid in a larger range of scales. In particular, they showed that a certain quantity, namely the propagator defined in eq.~(\ref{propd}), which measures the sensitivity of density and velocity perturbations to a variation in their initial conditions, can be computed {\it exactly}, {\em i.e.} at all orders in PT, in the large wavevector $k$ (small scale) limit. CS finding prompted a certain number of independent investigations on possible ways to resum PT contributions at all orders, both for the propagator \cite{Anselmi:2010fs, BV08, Bernardeau:2008fa, Skovbo:2011xq}, and for the directly measurable PS \cite{Pietroni08, MP07b, Taruya2007, Matsubara07, 2011MNRAS.416.1703E, Wang:2011fj, Wang:2012fr, Juergens:2012fk}. An alternative method, having the Zeldovich approximation as starting point, was proposed in \cite{Tassev:2012cq, Tassev:2011ac}. The status of these methods to date can be summarized as follows: at $z \agt 1$ the PS can be computed at a few percent accuracy (in comparison with state of the art N-body simulations) in the BAO range of scales ($0.05\alt k\alt 0.25\,\mathrm{h/Mpc}$), the accuracy degrading quite rapidly at higher wave numbers (smaller scales) and smaller redshifts. Moreover, the computational time of these approaches, though much smaller than for N-body simulations, is still in the few hours range for a single PS, thereby making the implementations of MCMC's still quite problematic.

\begin{figure}
\centerline{\includegraphics[width = 15cm,keepaspectratio=true]{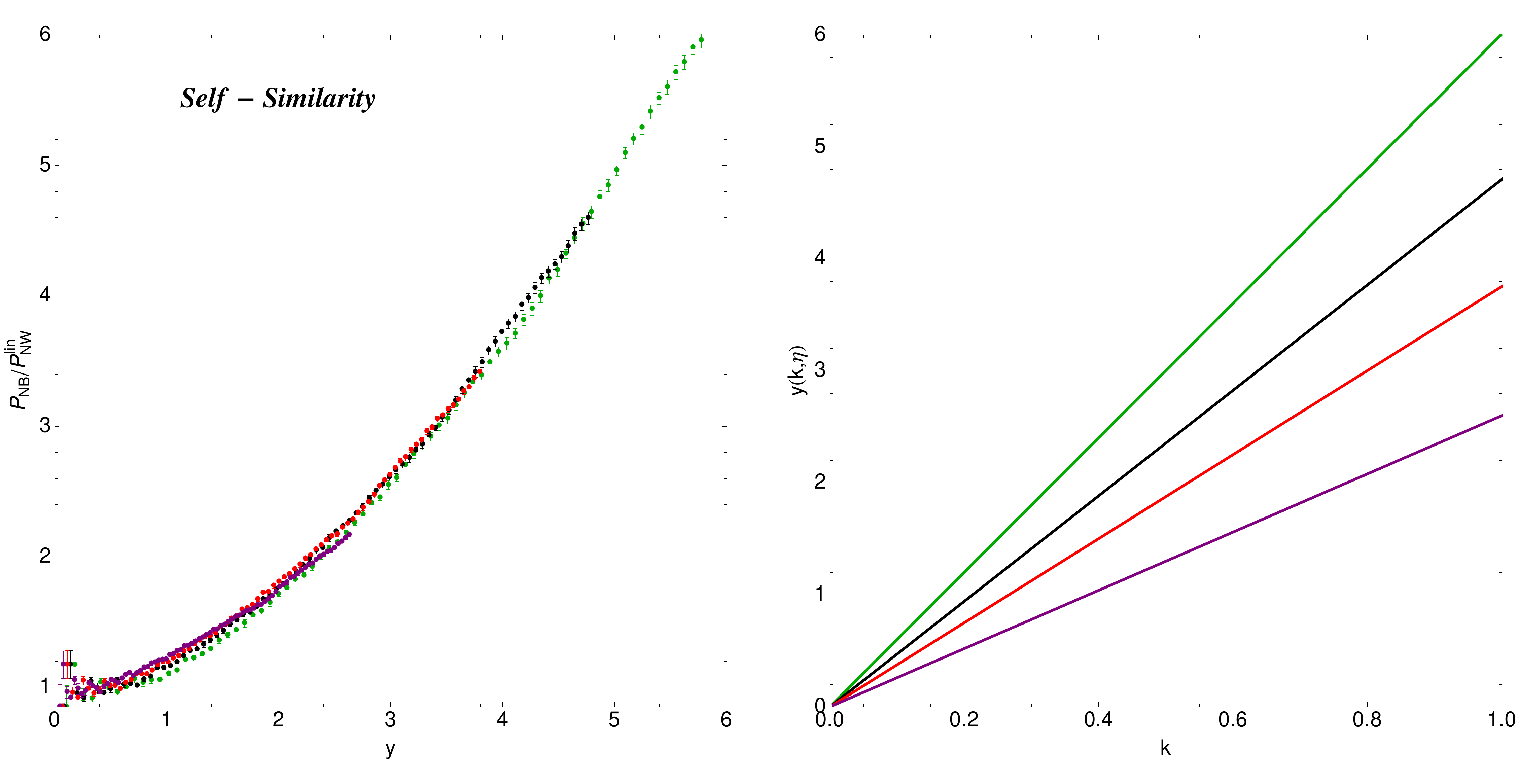}}
\caption{(left) The nonlinear PS from the N-body simulations of \cite{Sato:2011qr}, divided by the no-wiggle PS of \cite{eisensteinhu}, plotted against the variable $y$ defined in eq.~\re{ly}. The color-code is the following: green for $z=0$,  black for $z=0.5$, red for $z=1$, and purple for $z=2$. Each PS has been truncated at $k=1\,\mathrm{h/Mpc}$. (right) The relation between $y$ and $k$ at different redshifts.}
\label{selfsim}
\end{figure}

In this paper we present a new computational scheme which overcomes the present limitations of resummed PT approaches in both respects: it greatly enlarges the range of scales in which it gives results accurate at the percent level, and it greatly reduces computational times. At $z=1$ we can compute the PS up to $k\simeq 1\,\mathrm{h\;Mpc^{-1}}$ at the percent level, in a time comparable to that of a simple 1-loop computation, namely, $O(1)$ minutes. This opens the road both to parameter estimation via MCMCs, and to the extension of these methods from the BAO physics to weak lensing measurements. 

At the technical level, our main result is a resummation of the dominant PT corrections to the PS in the range of scales mentioned above. In the CS resummation for the propagator the effective expansion parameter in the large $k$ limit turns out to be \cite{RPTb}
\beq
y\equiv e^{\eta} \sigma_v k\,,
\label{ly}
\eeq
where $\eta=\log D(z)/D(z_{in})$ ($D(z)$ being the linear growth factor), and 
\beq
\sigma_v^2\equiv \frac{1}{3}\int d^3q \frac{P^0(q)}{q^2}\,,
\label{sv}
\eeq
with $P^0(q)$ the linear PS evaluated at the initial redshift,  to be formally sent to infinity (in practice, in this paper we will use $z_{in} = 100$). 
Our starting point is the empirical realization that, at wave numbers $k$ larger than the BAO range, the nonlinear corrections to the PS take the form of a multiplicative function of the variable $y$ defined in  eq.~\re{ly}. In fig.~\ref{selfsim} we plot the ratio between the nonlinear PS from the N-body simulations of \cite{Sato:2011qr} at different redshifts, and  the smooth linear PS by Eisenstein and Hu \cite{eisensteinhu}, as a function of  $y$.  It is clear that, for  $y> O(1)$, all the ratios are well approximated by an universal function of $y$. Therefore our goal is to identify, at each order in PT, the leading large $y$ corrections which, once resummed at all orders, give the $y$-function plotted in fig.~\ref{selfsim}. In the range of scales considered in this paper ($k\alt 1 \,\mathrm{h\;Mpc^{-1}}$)
it is given, as a first approximation,  by the simple analytic expression in eq.~\re{approxasy}, which is obtained as a limit of the more refined eq.~\re{TReik} discussed in sect.~\ref{largek}.

We formulate a very efficient way to reorganize the PT expansion, based on evolution equations in time which are exact at all orders in PT. A similar approach was already presented in ref.~\cite{Anselmi:2010fs} for the propagator, where it was used to reproduce the CS result, and also to include next-to-leading corrections. Here we derive the relevant equation for the PS and, after implementing an approximation analogous to the one leading to the CS result, we derive its behavior in the large $k$ or, better, large $y$, limit. At the same time, at low $k$, the equation is solved by the 1-loop PS, thereby providing an interpolation between the two correct behaviors in the two extremes of the physically interesting range of scales. We should warn the reader here, as we will repeat again, that the range of scales where the eulerian approach is applicable is limited in the UV (small scales) by multi streaming, {\it i.e.} by small scale velocity dispersion, a physical effect which is absent from the approach even at the non-perturbative level. Nevertheless, the very large $k$ limit or, more precisely, the limit in which all mode coupling is only with wave numbers $q$ much lower than the scale $k$ one is interested in, can be discussed exactly, and provides an useful starting point for the description of the intermediate scales, where the Eulerian approach is expected to work. 

The evolution equation contains two kinds of kernel functions. One is the same entering the equation for the propagator, and was already discussed in ref.~\cite{Anselmi:2010fs}. A computation of this kernel function at 1-loop provides, by virtue of our evolution equation, the CS propagator at all loop order, as well as part of the leading corrections to the PS. The resummation of the leading corrections of the remaining part, which is responsible for mode-mode coupling, is the main result of this paper.

As we have already anticipated, our combined resummation and interpolation procedure works remarkably well: its performance degrades only for (low) redshifts and (small) scales where the effect of multi-streaming, the intrinsic limit of eulerian PT, is known to become relevant \cite{Pueblas:2008uv,Valageas:2010rx}. In other words, the comparison with high resolution N-body simulations presented in this work shows that, at the few percent level, our approach reproduces well the physics contained in the Euler-Poisson system, on which eulerian PT is founded. As a consequence, it provides the best starting base for methods aimed at going beyond the single-stream approximation, as that proposed in \cite{Pietroni:2011iz}.

The paper is organized as follows. In sec.~\ref{EOM} we review eulerian PT in the compact form discussed in \cite{RPTb, MP07b}, the diagrammatic approach to the PT expansion, and we recall the exact expressions for the fully nonlinear propagator and PS, which provide the starting points for their evolution equations. In sec.~\ref{TRGEq} we derive the evolution equations for the propagator and for the PS. In sec.~\ref{1loop} we show how, in the $k\to 0$ limit, the PS equation is solved by the standard 1-loop expression for the PS, and in sec.~\ref{smallk}  we discuss how a simple upgrading of the 1-loop expression provides an all-order resummation valid in the small $k$ regime. Then, in sec.~\ref{largek} we discuss the large $k$ limit and discuss how to interpolate between this and the low $k$ regime.  In sec.~\ref{results} we present our numerical results and their comparison with N-body simulations. Finally, in sec. \ref{discussion} we discuss our results and possible developments.

In \ref{eRPT} we discuss an alternative framework to do RPT, and in \ref{3L} we prove an useful relation valid in the large $k$ regime. In \ref{AA} and \ref{AB} we give all the explicit formulas needed for the numerical implementation of the method.

\section{Nonlinear fluid equations and perturbation theory}
\label{EOM}
Eulerian PT \cite{PT} aims at solving the system of three fluid equations: continuity, Euler, Poisson. Starting from the case of Einstein-de Sitter cosmology, the equations can be written as follows
\beqra
&&\frac{\partial\,\delta_m}{\partial\,\tau}+
{\bf \nabla}\cdot\left[(1+\delta_m) {\bf v} \right]=0\,,\nonumber\\
&& \frac{\partial\,{\bf v}}{\partial\,\tau}+{\cal H}\,{\bf v} \, + ( {\bf v} 
\cdot {\bf \nabla})  {\bf v}= - {\bf \nabla} \phi\,,\nonumber\\
&&\nabla^2 \phi = \frac{3}{2}\,\,{\cal H}^2  \, \, \delta_m\, ,
\label{Euler}
\eeqra
where  ${\cal H}= d \log a/d \tau$ is the Hubble Parameter in conformal time,  while $\delta_m({\bf x},\,\tau)$ and  ${\bf v} ({\bf x},\,\tau)$ are the DM number-density fluctuation and the DM peculiar velocity field, respectively.

Defining, as usual, the velocity divergence $\theta(\bx,\,\tau) = 
\nabla \cdot {\bf v} (\bx, \,\tau)$, and going to
Fourier space, the equations in (\ref{Euler}) can be expressed as
\beqra
&&\frac{\partial\,\delta_m({\bf k}, \tau)}{\partial\,\tau}+\theta({\bf k}, 
\tau) \nonumber\\
&& \qquad + \int d^3\bq\, d^3\bp \,\delta_D({\bf k}-\bq-\bp)
 \alpha(\bq,\bp)\theta(\bq, \tau)\delta_m(\bp, \tau)=0\,,\nonumber\\
&&\frac{\partial\,\theta({\bf k}, \tau)}{\partial\,\tau}+
{\cal H}\,\theta({\bf k}, \tau)  +\frac{3}{2} {\cal H}^2 \,
\delta_m({\bf k}, \tau)\nonumber\\
&& \qquad   +\int d^3\bq \,d^3\bp \,\delta_D({\bf k}-\bq-\bp) 
\beta(\bq,\bp)\theta(\bq, \tau)\theta(\bp, \tau) = 0 \,.\label{EulerFourier}
\eeqra
The nonlinearity and non-locality of the fluid equation are encoded in 
the two functions
\beq\alpha(\bq,\bp )= \frac{(\bp + \bq) \cdot \bq}{q^2}\,,\quad \quad
\beta(\bq,\bp ) = \frac{(\bp + \bq)^2 \,
\bp \cdot \bq}{2 \,p ^2 q^2}\,,
\eeq
which couple different modes of density and velocity fluctuations. 

One can write Eqs.~(\ref{EulerFourier}) in a compact form \cite{RPTb}. First, 
we introduce the doublet $\vp_a$ ($a=1,2$), given by
\beq\left(\begin{array}{c}
\varphi_1 ( {\bf k}, \eta)\\
\varphi_2 ( {\bf k}, \eta)  
\end{array}\right)
\equiv 
e^{-\eta} \left( \begin{array}{c}
\delta_m  ( {\bf k}, \eta) \\
-\theta  ( {\bf k}, \eta)/{\cal H}
\end{array}
\right)\,,
\label{doppietto}
\eeq
where the time variable has been replaced by the logarithm of 
the scale factor,
\[ 
\eta= \log\frac{a}{a_{in}}\,,
\]
$a_{in}$ being the scale factor at a conveniently remote epoch, 
in which all the relevant scales are well inside the linear regime. 

Then, we define a {\it vertex} function, 
$\gamma_{abc}({\bf k},{\bf p},{\bf q}) $ ($a,b,c,=1,2$) 
whose only independent, non-vanishing, elements are
\beqra
&&\gamma_{121}({\bf k},\,{\bf p},\,{\bf q}) = 
\frac{1}{2} \,\delta_D ({\bf k}+{\bf p}+{\bf q})\, 
\alpha(\bp,\bq)\,,\nonumber\\
&&\gamma_{222}({\bf k},\,{\bf p},\,{\bf q}) = 
\delta_D ({\bf k}+{\bf p}+{\bf q})\, \beta(\bp,\bq)\,,
\label{vertice}
\eeqra
and 
$\gamma_{121}({\bf k},\,{\bf p},\,{\bf q})  = 
\gamma_{112}({\bf k},\,{\bf q},\,{\bf p}) $.

The two equations (\ref{EulerFourier}) can now be rewritten in a compact form as
\beq
\partial_\eta\,\varphi_a({\bf k}, \eta)= -\Omega_{ab}\,
\varphi_b({\bf k}, \eta) + e^\eta 
\gamma_{abc}({\bf k},\,-{\bf p},\,-{\bf q})  
\varphi_b({\bf p}, \eta )\,\varphi_c({\bf q}, \eta ),
\label{compact}
\eeq
where 
\beq 
\Omega= \left(\begin{array}{rr}
\ds 1 & \ds -1\\&\\
\ds -\frac{3}{2} & \ds \frac{3}{2} \end{array}
\right)\,.
\label{bigomega}
\eeq
Repeated indices are summed over, and integration over momenta $\bq$ and $\bp$ is understood.\\

To extend the validity of this approach to $\Lambda$CDM, we will reinterpret the variable $\eta$ as the logarithm of the linear growth factor of the growing mode, i.e. ~\cite{PT,RPTb,Pietroni08},
\beq
\eta=\ln (D/D_{in})\,,
\eeq and we redefine the field in Eq.~(\ref{doppietto}) as 
\beq 
\left(\begin{array}{c}
\varphi_1 ( {\bf k}, \eta)\\
\varphi_2 ( {\bf k}, \eta)  
\end{array}\right)
\equiv 
e^{-\eta} \left( \begin{array}{c}
\delta_m  ( {\bf k}, \eta) \\
-\theta  ( {\bf k}, \eta)/{\cal H }f
\end{array}\right)\,,
\label{PhiF}
\eeq
with $f=d \ln D/d\ln a$. As discussed in ~\cite{Pietroni08}, the above approximation is accurate at better than $1\%$ level in the whole range of redshifts and scales we are interested in. 

If we consider the linear equations (obtained in the $e^{\eta}\gamma_{abc}\rightarrow 0$ limit) we can define the {\it linear retarded propagator}  as the operator giving the evolution of the field $\varphi_{a}$ from $\eta_{in}$ to $\eta$,
\beq
\vp^{0}_{a}(\bk,\eta)=g_{ab}(\eta,\eta_{in})\vp^{0}_{b}(\bk,\eta_{in})\,,
\eeq
where the $``0"$ index stands for the linear approximation to the full solution.
The linear propagator obeys the equation 
\beq
(\delta_{ab}\partial_{\eta}+\Omega_{ab})g_{bc}(\eta,\eta_{in})=\delta_{ac}\delta_D(\eta-\eta_{in}).
\eeq
with causal boundary conditions. It is given explicitly by the following expression \cite{RPTb},
\beq 
g_{ab}(\eta,\eta^\prime) =\left[ {\bf B} + {\bf A}\, e^{-5/2 
(\eta -\eta^\prime)}\right]_{ab}\, \theta(\eta-\eta^\prime)\,,
\label{proplin}
\eeq
 with $\theta$ the 
step-function, and
\beq {\bf B} = \frac{1}{5}\left(\begin{array}{cc}
3 & 2\\
3 & 2
\end{array}\right)\,\qquad {\mathrm{and}} \qquad
{\bf A} = \frac{1}{5}\left(\begin{array}{rr}
2 & -2\\
-3 & 3
\end{array}\right)\,.
\eeq
The growing ($\vp_a \propto \mathrm{const.}$) and the decaying 
($\vp _a\propto \exp(-5/2( \eta - \eta^\prime))$) modes can be selected by 
considering initial fields $\vp_a$ proportional to 
\beq u_a = \left(\begin{array}{c} 1\\ 
1\end{array}\right)\,\qquad\mathrm{and} 
\qquad v_a=\left(\begin{array}{c} 1\\ -3/2\end{array}\right)\,,
\label{ic}
\eeq
respectively.

\begin{figure}
\centerline{\includegraphics[width = 11cm,keepaspectratio=true]{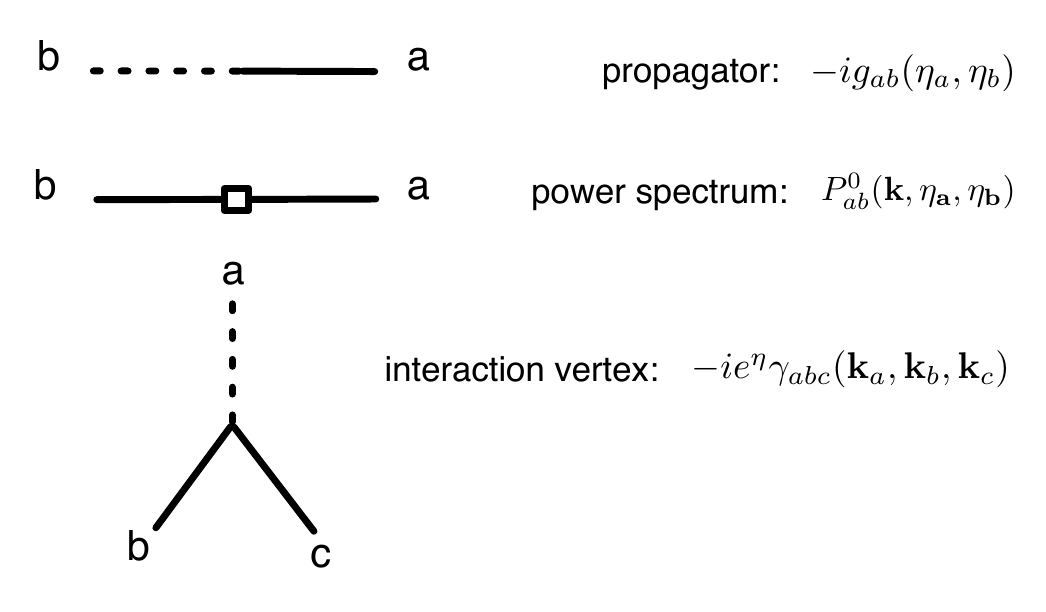}}
\caption{The Feynman Rules for cosmological perturbation theory}
\label{FEYNMAN}
\end{figure}
\begin{figure}
\centerline{\includegraphics[width = 11cm,keepaspectratio=true]{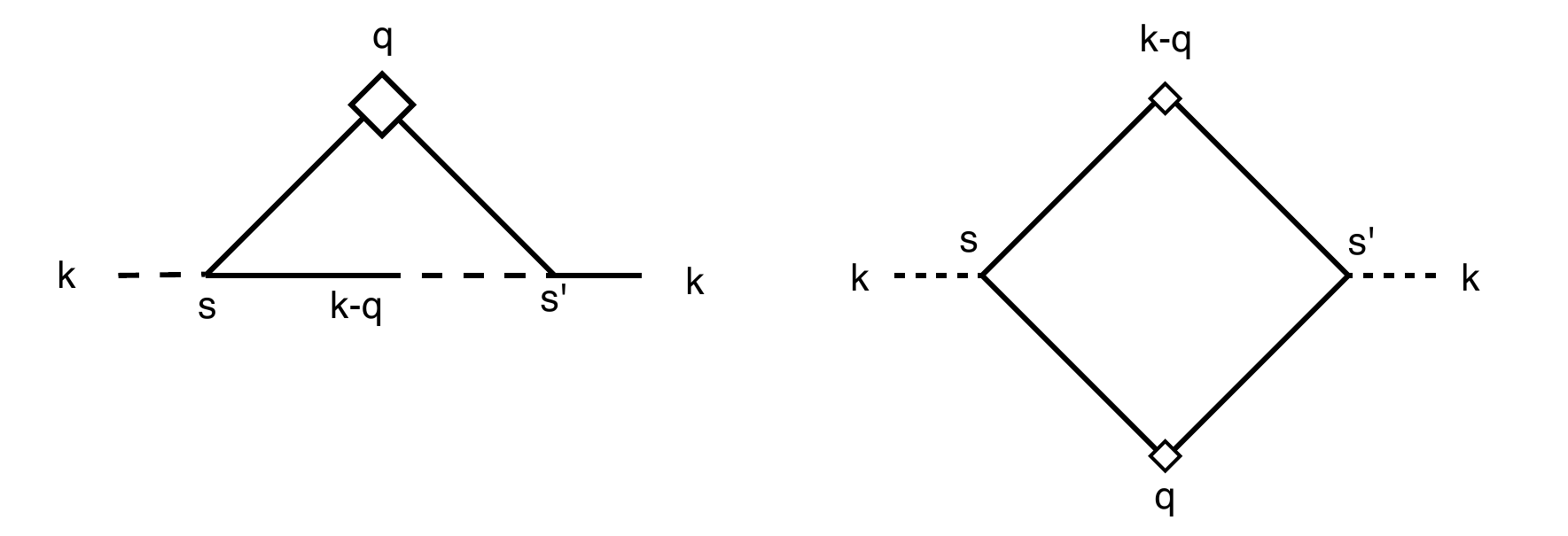}}
\caption{The two 1PI functions $\Sigma^{(1)}_{ab}$ and $\Phi^{(1)}_{ab}$ at 1-loop, given explicitly by eqs.~\re{SIGMA1L} and \re{phicomp}.}
\label{Phi1}
\end{figure}

In the following, we will be interested in the PS 
\beq
\langle \vp_a(\bk,\eta)\vp_b(\bk^\prime,\eta^\prime) \rangle
 \equiv \delta_D(\bk +\bk^\prime) P_{ab}(k; \eta,\eta^\prime)\,,
 \label{PSd}
\eeq
and in the propagator,
\beq
\langle \frac{\delta \vp_a(\bk,\eta)}{\delta \vp_b(\bk^\prime,\etap)} \rangle
\equiv i\, \delta_D(\bk +\bk^\prime) G_{ab}(k;\eta,\eta^\prime)\,,
 \label{propd}
\eeq
which gives the response of the field  at time $\eta$ to a change in the field at an earlier time $\etap <\eta$.
The linear approximations of the fully nonlinear quantities above are given by the linear PS
\beq
P^0_{ab}(k; \eta,\etap)=g_{ac}(\eta,\eta_{in})g_{bd}(\eta,\eta_{in})P^{0}_{cd}(k;\etain,\etain) \,,
\label{linps}
\eeq
and by the linear propagator $g_{ab}(\eta,\etap)$ of eq.~(\ref{proplin}), respectively. 

These quantities, as well as all higher order correlation functions, can be computed perturbatively by solving iteratively the nonlinear equations (\ref{compact}), supplemented by the initial conditions at the time $\etain$ for the PS, the bispectrum, and, in principle, all higher correlation functions. In this paper, we will limit ourselves to the case of gaussian initial conditions, and therefore we will assume that the initial conditions are fully characterized by the PS only. 

A very convenient way to organize the perturbative expansion is to use the diagrammatic language discussed in \cite{RPTa} and in \cite{MP07b} (for details on the derivation of formalism used in this paper, the reader is referred to the latter). The building blocks are given in fig.~\ref{FEYNMAN}, and are the linear propagator, the linear PS, and the interaction vertex of eq.~(\ref{vertice}). Propagators and PS can only be joined according to the rule encoded in the vertex, {\it i.e.}, two continuous ends join with a dashed one. Notice, moreover, that according to eq.~(\ref{linps}), a linear PS can be seen as the initial PS, represented by the empty square in fig.~\ref{FEYNMAN}, sandwiched between two linear propagators. This will be useful in the following. 

The full nonlinear PS and propagator have the following structures \cite{MP07b}

\beqra
P_{ab}(k; \eta,\etap) &=& G_{ac}(k;\eta,\etain)
G_{bd}(k;\etap,\etain) P_{cd}(k; \etain,\etain)\,\nonumber\\
&&+
 \int d s\,
d s^\prime\;
G_{ac}(k;\eta,s)
G_{bd}(k;\etap, s^\prime) 
\Phi_{cd}(k; s, s^\prime)\,,\nonumber\\&&
\label{fullP}
\eeqra
and
\beq
 G_{ab} (k; \eta, \etap) = \left[g^{-1}_{ba} - 
\Sigma_{ba}\right]^{-1}(k; \eta, \etap)\,,
\label{fullG}
\eeq
where the last expression has to be interpreted in a formal sense, that is,
\beqra
 &&G_{ab} (k;\, \eta, \etap)=g_{ab} (\eta-\etap) \nonumber\\
 &&\qquad\qquad+ \int d s\,
d s^\prime\; g_{ac} (\eta-s) \Sigma_{cd}(k;\, s, s^\prime) g_{db} (s^\prime - \etap) + \cdots\,.
\label{GEXPANSION}
\eeqra
Notice that, both in (\ref{fullP}) and in (\ref{fullG}) and everywhere else unless explicitly stated otherwise, the time integrals run from $-\infty$ to $+\infty$. However, in practice, due to the causal properties of the retarded propagators (see eq.~(\ref{proplin})), the range of time integrations turns out to be always finite.

Eqs.~(\ref{fullP}) and (\ref{fullG}) can be obtained either by analyzing the diagrammatic structure of the contributions at an arbitrarily high order, as in \cite{RPTa}, or by functional methods, as in \cite{MP07b}. They are exact relations ({\it i.e.} valid at all orders in PT), and are valid both for gaussian and for non-gaussian initial conditions. Eq.~(\ref{fullG}) has been already exploited as the starting point of the derivation of the evolution equation for the propagator discussed in \cite{Anselmi:2010fs}. In the next section we will derive the evolution equation for the PS, starting from eq.~(\ref{fullP}).

In the following, we will take the initial conditions for the PS, $P_{cd}(k; \etain,\etain)$, to be given by the linear PS in the growing mode, $P^0(k)u_c u_d$, which is equivalent to neglecting any nonlinear effect in the initial PS on the range of scale of interest. This is of course exact in the $\etain\to -\infty$ limit. The linear PS in eq.~(\ref{linps}) then becomes  $P^{0}_{ab}(k;\eta,\etap)=u_a u_b P^0(k)$, where we have used the property of the linear propagator, $g_{ac}(\eta,\etap)u_c = u_a$, which can be checked using eqs.~(\ref{proplin}) and (\ref{ic})\footnote{Remember that our fluctuations, defined in eq.~(\ref{doppietto}), are constant at the linear level.}.

The crucial quantities entering eqs.~(\ref{fullP}) and (\ref{fullG}) above are the  1-particle-irreducible (1PI) functions $\Sigma_{ab}$ and $\Phi_{ab}$, where 1PI means, as usual in Feynman diagrammatic language, that the diagrams contributing to these quantities cannot be separated into two disjoint parts by cutting a single PS or propagator line. $\Sigma_{ab}(k;s,s')$ connects a dashed end at time $s$ to a continuous one at time $s'<s$, whereas $\Phi_{ab}(k;s,s')$, connects two continuous lines at $s$ and $s'$ with no time ordering. The lowest order contributions to these 1PI functions are represented by the diagrams in fig.~\ref{Phi1}, and correspond to the explicit formulae given in \ref{AA}.

The information contained in the 1PI functions $\Sigma_{ab}$ and $\Phi_{ab}$ is completely equivalent to that encoded in the connected ones, namely the propagator and the 
PS. However, working with the 1PI functions inside the time evolution equations to be introduced in next section presents some advantage, already exploited in \cite{Anselmi:2010fs}. Indeed,  this approach provides a natural guide to interpolate between the $k\to0$ and $k\to \infty$ limits  and, moreover,  it is more suitable to be used for cosmology with time and/or scale dependent linear growth factors, which can be encoded in a time and scale dependence on the $\Omega_{ab}$ matrices \re{bigomega}, see \cite{Pietroni08}.

The main focus of this paper  will be on the nonlinear PS evaluated at equal times $\eta=\etap$ (hereafter, the PS), that is,
\beq
P_{ab}(k; \eta) \equiv P_{ab}(k; \eta,\eta)\,.
\eeq
\section{The evolution equations}
\label{TRGEq}
In order to derive the time-evolution of $P_{ab}(k; \eta)$ we need that of $G_{ab} (k; \eta, \etap)$, which can be obtained by using the property of the linear propagator
\beq
\partial_\eta\, g_{ab} (\eta, \etap) =\delta_{ab}\, \delta_D(\eta-\etap) - \Omega_{ac} \,g_{cb} (\eta , \etap) \,,
\label{dglin}
\eeq
in eq.~(\ref{GEXPANSION}). We get (see \cite{Anselmi:2010fs}),
\beq
\partial_\eta\, G_{ab} (k;\, \eta, \etap) = \delta_{ab}\, \delta_D(\eta-\etap) - \Omega_{ac} \,G_{cb} (k;\, \eta, \etap)+\Delta G_{ab}(k;\,\eta,\etap)\,,
\label{dgtot}
\eeq
where
\beq
\Delta G_{ab}(k;\,\eta,\etap)\equiv \int ds^\prime \; \Sigma_{ad}(k;\, \eta, s^\prime)\,G_{db} (k;\, s^\prime, \etap)\,.
\label{deltaG}
\eeq
Notice that the terms containing the Dirac delta in (\ref{dglin}) and (\ref{dgtot}) where not written in \cite{Anselmi:2010fs}, because in that paper we always considered propagators at different times. Here, on the other hand, those terms are important when the propagator is inside a time integral, as in the second line of eq.~(\ref{fullP}).

Taking the $\eta$ derivative of $P_{ab}(k; \eta)$ as given by (\ref{fullP}), we get
\beqra
&&\partial_\eta \,P_{ab}(k; \eta) = -\Omega_{ac} \,P_{cb}(k; \eta)  -\Omega_{bc} \,P_{ac}(k; \eta)\nonumber \\
&&\;\; +\Big(\Delta G_{ac}(k;\eta,\etain)  G_{bd }(k;\eta,\etain)+  G_{ac}(k;\eta,\etain)  \Delta G_{bd }(k;\eta,\etain)\Big) P^0(k) u_c u_d \nonumber\\
&&\;\;+ \int \, ds'\;\big[\Phi_{ac}(k;\eta,s') G_{bc}(k;\,\eta,s')+ G_{ac}(k;\,\eta,s')\Phi_{cb}(k;s',\eta) \big]\,\nonumber\\
&&+ \int \, ds\, ds' \;\Phi_{cd}(k;s,s') \Big(\Delta G_{ac}(k;\eta,s) G_{bd}(k;\,\eta,s')\nonumber\\
&&\qquad\qquad\qquad\qquad+  G_{ac}(k;\eta,s) \Delta G_{bd}(k;\,\eta,s') \Big)\,,\nonumber\\
\label{Texact}
\eeqra
which is of course still an exact equation. 

In  \cite{Anselmi:2010fs}, we showed that in the large $k$ limit $\Delta G_{ac}$  takes the form
\beq
\Delta G_{ac}(k;\,\eta,s) \simeq H_{{\bf a}}(k;\, \eta,s)\, G_{{\bf a}c}(k;\,\eta,s)\, ,
\label{TRGL}
\eeq
where
\[
H_{{ a}}(k;\, \eta,s) \equiv  \int_{s}^{\eta} d s''\, \Sigma_{{ a}e}^{(1)}( k;\,\eta\,,s'')\,u_e\,,
\]
with $\Sigma_{{ a}d}^{(1)}$ the 1-loop approximation to the full $\Sigma_{ad}$, see eqs.~\re{sigform} and \re{SIGMA1L},
and the boldface index indicates that we are not summing over that index even if it is repeated. The same factorization, eq.~\re{TRGL}, holds at $k\to 0$ where the 1-loop PS is obtained from the solution of the evolution equation. Therefore, in  \cite{Anselmi:2010fs}, a natural way to interpolate between the two extreme limits has been identified in using eq.~\re{TRGL} also for intermediate $k$'s.

We will denote the solution of eq.~(\ref{dgtot}) with the approximation (\ref{TRGL}) as $\bar G_{ab}(k;\,\eta,\etap)$, which we will use in \re{Texact}. As it was discussed thoroughly in  \cite{Anselmi:2010fs}, the solution $\bar G_{ab}(k;\,\eta,\etap)$ is exact both in the low and in the large $k$ limits. At low $k$ it reproduces the 1-loop propagator
\beqra
&&\bar G_{ab}(k;\,\eta,\etap) \to g_{ab}(\eta,\etap) \nonumber\\
&& \qquad \quad+\int ds ds'\, g_{ac}(\eta,s) \Sigma_{cd}^{(1)}(k;s,s') g_{db}(s',\etap)\, (\mathrm{for}\;\;k\to0)\,,
\label{G1loop}
\eeqra
whereas at large $k$ it reproduces the exact result obtained by CS in \cite{RPTb}
\beqra
&&\bar G_{ab}(k;\,\eta,\etap) \to G_{ab}^{eik}(k;\,\eta,\etap) \equiv g_{ab}(\eta-\etap) \exp\left[-k^2\sigma_v^2 \frac{(e^\eta-e^{\etap})^2}{2}\right],\nonumber\\
&&\qquad\qquad\qquad\qquad\qquad (\mathrm{for}\;\;k\to\infty)\,,
\label{resCS}
\eeqra
where $\sigma_v^2$ is defined in eq.~\re{sv}, and ``$eik$'' stands for the {\it eikonal} limit, in which the above expression is exact \cite{Bernardeau:2011vy, Bernardeau:2012aq}, see Sect.~\ref{largek}.

\section{Recovering 1-loop}
\label{1loop}
As a first attack to a practical and fast solution of eq.~(\ref{Texact}), we identify the limit  in which it reproduces the 1-loop result. It is obtained by using eq.~\re{TRGL} in
\re{Texact} and by setting 
\beq
 G_{ab} \to g_{ab}\,,\qquad \Phi_{ab}\to \Phi^{(1)}_{ab}\,,
\label{2app}
\eeq
in the second and third lines, where $\Phi^{(1)}_{ab}$ is  the 1-loop approximation to $\Phi_{ab}$. Moreover, the fourth and fifth lines, containing at least  2-loop order quantities, are consistently neglected.
Thus, we get
\beqra
\partial_\eta \,P_{ab}^{(1)}(k; \eta) &=& -\Omega_{ac} \,P_{cb}^{(1)}(k; \eta)  -\Omega_{bc} \,P_{ac}^{(1)}(k; \eta)\nonumber \\
&& + P^0(k)\Big(H_{a}(k;\, \eta,\etain)u_b +H_{b}(k;\, \eta,\etain)u_a\Big) \nonumber\\
&&+ \int  \, ds\;\Big(\Phi_{ad}^{(1)}(k;\eta,s) g_{bd}(\eta,s)+ g_{ad}(\eta,s)\Phi_{db}^{(1)}(k;s,\eta) \Big)\,. \nonumber\\
\label{T1loop}
\eeqra

The solution of the equation above exactly reproduces the 1-loop PS, 
\beqra
P_{ab}^{(1)}(k;\eta) &=& P^0(k) \Big[ u_a u_b \nonumber\\
&&+ \int ds \left( g_{ae}(\eta,s) u_b + g_{be}(\eta,s) u_a\right) H_e(k;s,\etain)\Big] \nonumber\\
&&+\int ds \,ds^\prime g_{ac}(\eta,s) g_{bd}(\eta,s^\prime)\Phi_{cd}^{(1)}(k;s,s^\prime)\,.
\label{PS1loop}
\eeqra
as can be checked directly by taking the $\eta$-derivative of the expression above.

\section{The small $k$ limit}
\label{smallk}
The approximations leading to the 1-loop result suggest the first step to take to go beyond, and to obtain a first resummation, containing infinite orders of the PT expansion. Indeed, by promoting the linear PS appearing in eq.~\re{T1loop} to the nonlinear and time-depedent one,  we get
\beqra
\partial_\eta \,P_{ab}^{R1}(k; \eta) &=& -\Omega_{ac} \,P_{cb}^{R1}(k; \eta)  -\Omega_{bc} \,P_{ac}^{R1}(k; \eta)\nonumber \\
&& + H_{{\bf a}}(k;\, \eta,\etain)\,P_{{\bf a}b}^{R1}(k; \eta) +H_{{\bf b}}(k;\, \eta,\etain)\,P_{a{\bf b}}^{R1}(k; \eta) \nonumber\\
&&+ \int \, ds\;\big[\Phi_{ad}^{(1)}(k;\eta,s) g_{bd}(\eta,s)+ g_{ad}(\eta,s)\Phi_{db}^{(1)}(k;s,\eta) \big]\,. \nonumber\\
\label{TR1}
\eeqra
To understand what this approximation corresponds to, it is instructive to set to zero the third line of the above equation. Then, the equation can be solved exactly, to get
\beq
\left.P_{ab}^{R1}(k; \eta)\right|_{\Phi^{(1)}_{ab}=0}= \bar G_{ac}(k;\,\eta,\etain)\bar G_{bd}(k;\,\eta,\etain)u_c u_d P^{0}(k)\,.
\label{GGP}
\eeq
The above expression contains infinite orders in PT, and corresponds to the linear PS multiplied by two renormalized propagators. By turning $\Phi^{(1)}_{ab}$ on we are adding perturbatively the effect of mode-mode coupling.

A step further, which does not increase the computing time too much, consists in improving the approximation in eq.~(\ref{2app}), by using the 1-loop approximation for $ G_{ab}$, instead of the linear one, in the second line of eq.~(\ref{TR1}), to get
 \beqra
\partial_\eta \,P_{ab}^{R2}(k; \eta) &=& -\Omega_{ac} \,P_{cb}^{R2}(k; \eta)  -\Omega_{bc} \,P_{ac}^{R2}(k; \eta)\nonumber \\
&& + H_{{\bf a}}(k;\, \eta,\etain)\,P_{{\bf a}b}^{R2}(k; \eta) +H_{{\bf b}}(k;\, \eta,\etain)\,P_{a{\bf b}}^{R2}(k; \eta) \nonumber\\
&&+ \int  \, ds\;\big[\Phi_{ad}^{(1)}(k;\eta,s) G_{bd}^{(1)}(k;\,\eta,s)+ G_{ad}^{(1)}(k;\,\eta,s)\Phi_{db}^{(1)}(k;s,\eta) \big]\,, \nonumber\\
\label{TR2}
\eeqra
where the 1-loop propagator has been given in eq.~(\ref{G1loop}).

Numerical results for the 1-loop approximation of eq.~\re{PS1loop} and for the small $k$ resummation discussed above, $P^{R2}_{ab}$, will be presented in Sect.~\ref{results}.

\section{The large $k$ limit}
\label{largek}
\subsection{The eikonal limit}

\begin{figure}
\centerline{\includegraphics[width = 13cm,keepaspectratio=true]{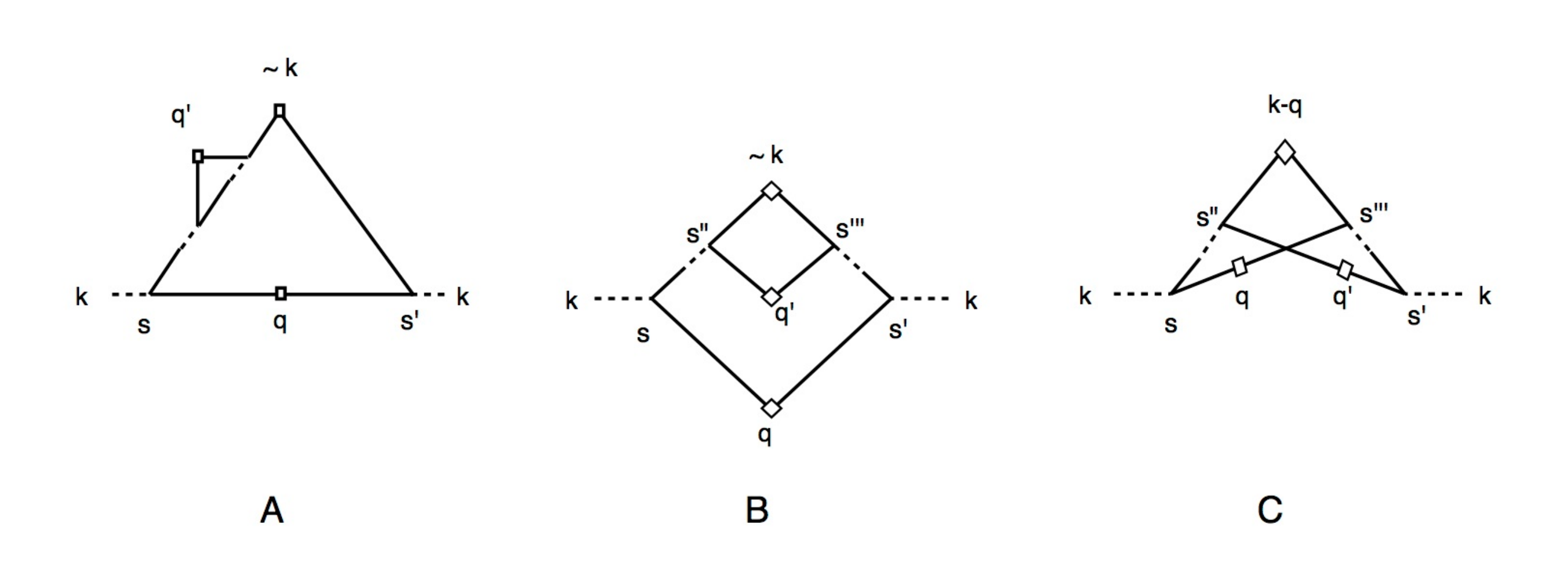}}
\caption{The 2-loop contributions to $\Phi_{ab}$ giving leading contributions in the large $k$ limit, with no vertex renormalization.}
\label{phi2loop}
\end{figure}

\begin{figure}
\centerline{\includegraphics[width = 8cm,keepaspectratio=true]{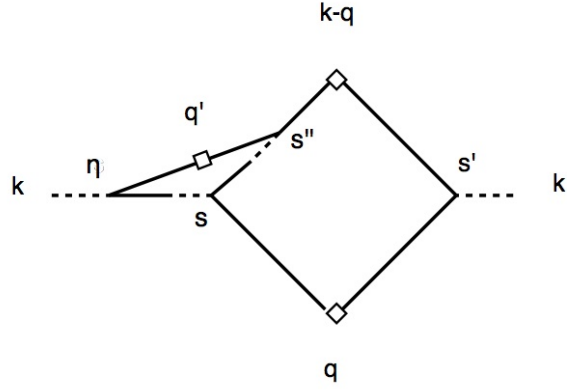}}
\caption{The dominant lowest order contribution to $\Phi_{ab}$ with a vertex renormalization.}
\label{nonphi}
\end{figure}

We will now derive the large-$k$ approximation of the equation for the PS, extending the derivation of ref.~\cite{Anselmi:2010fs} for the function $\Sigma_{ab}(k;\eta,\etap)$ to  the mode-coupling function, $\Phi_{ac}(k;\eta,\etap)$. In ref.~\cite{Anselmi:2010fs}, the inclusion of all the dominant configurations in the large $k$ limit allowed us to write eq.~\re{deltaG} as \re{TRGL}. In the present case, we will compute $\Phi_{ac}$ taking into account the diagram on the right of fig.~\ref{Phi1}, in the $k\gg q$ limit,  and all those diagrams at all orders,  that can be obtained from that one by attaching ``soft'' PS's, with momenta $q\ll k$ in all possible ways to the ``hard'' lines carrying momenta of $O(k)$, both inside the loop and as corrections to the vertices. For instance, at 2-loop order, the diagrams we will take into account are those of figs.~\ref{phi2loop} and \ref{nonphi}. We will denote the resulting $\Phi_{ac}(k;\eta,\etap)$ as $\Phi^{eik}_{ac}(k;\eta,\etap)$, since it is computed in the ``eikonal'' limit in which the wavenumber we are interested in,  $k$, is much larger than all the wave numbers it is effectively coupled to, see refs.~\cite{Bernardeau:2011vy, Bernardeau:2012aq}. In \ref{eRPT}  we discuss a framework to do renormalized PT, that we call eRPT, defined as a loop expansion around this extreme situation, where loop corrections progressively take into account the effects from intermediate modes, and restore the full momentum dependence at small and intermediate scales.

\begin{figure}
\centerline{\includegraphics[width = 10cm,keepaspectratio=true]{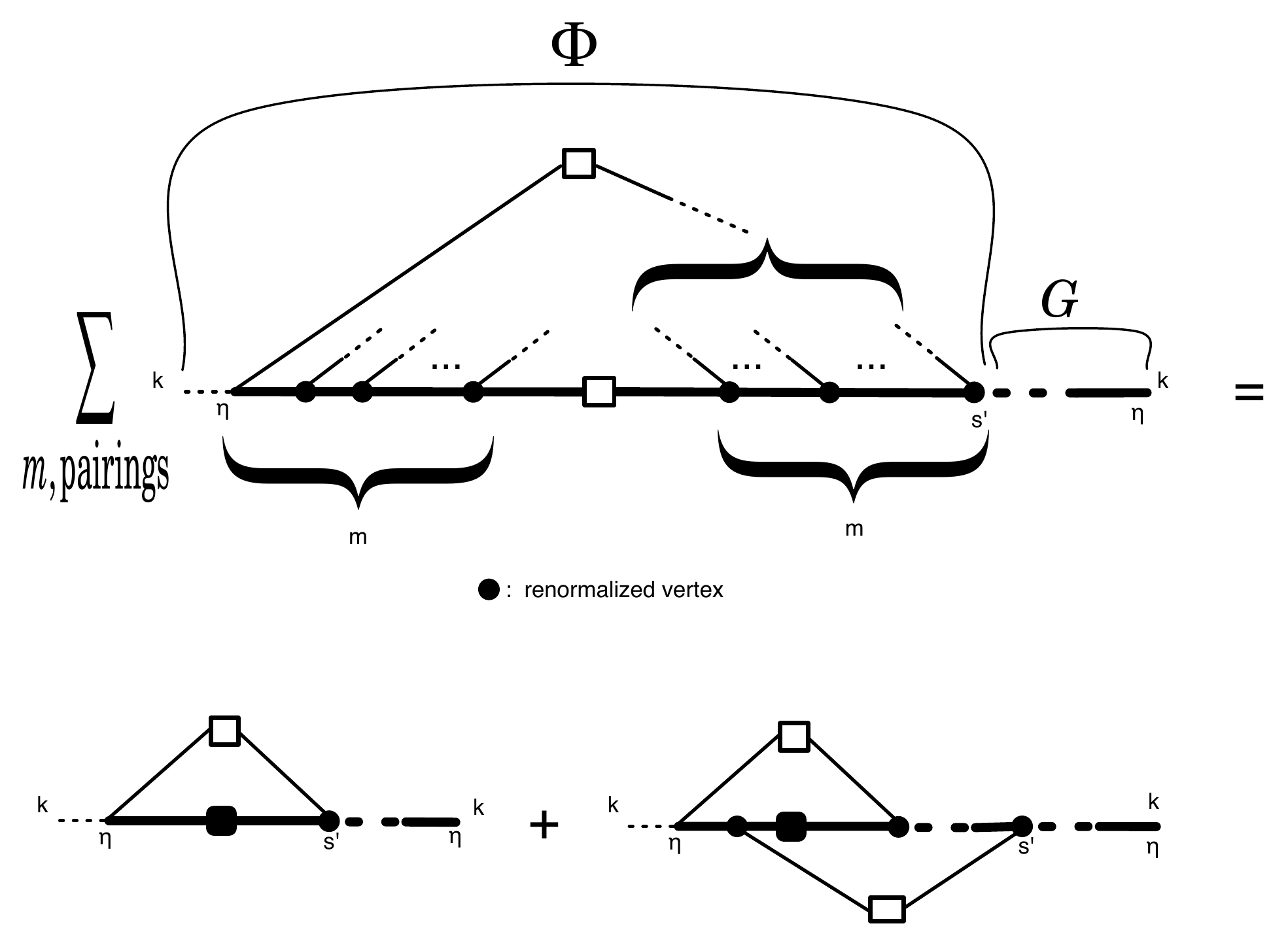}}
\caption{The diagrams contributing to $\Phi^{eik, L}_{ac}(k;\eta,s')  G_{bc}^{eik}(k;\eta,s')$. Thick lines and dots represent, respectively,  renormalized propagators and vertices in the eikonal limit.  The sum over $m$ and over all possible pairings of soft lines can be expressed as the two diagrams at the RHS. The black PS is given by eq.~\re{Pnle}. }
\label{PhiGeik}
\end{figure}

As we will show in \ref{3L}, in the eikonal limit the $n-$th order contributions (in standard PT)  to the last three lines of eq.~\re{Texact} can be written as

\beqra
&&\int \, ds\, ds' \;\sum_{l=1}^{n-1} \Big\{\Delta G^{eik,\,(l)}_{ac}(k;\eta,s) \big[ \Phi^{eik}_{cd}(k;s,s') G^{eik}_{bd}(k;\,\eta,s')\big]^{(n-l)}\nonumber\\
&&\qquad\qquad\quad+  \big[G^{eik}_{ac}(k;\eta,s)  \Phi^{eik}_{cd}(k;s,s') \big]^{(n-l)}\Delta G^{eik,\,(l)}_{bd}(k;\,\eta,s') \Big\} \nonumber\\
&&+ \int \, ds'\;\Big[\Phi^{eik}_{ac}(k;\eta,s') G^{eik}_{bc}(k;\,\eta,s')+ G^{eik}_{ac}(k;\,\eta,s')\Phi^{eik}_{cb}(k;s',\eta) \Big]^{(n)}\nonumber\\
&& = \int \, ds\, ds' \; \sum_{l=1}^{n-1} \Big\{  H^{eik}_{\bf a}(k;\,\eta,\etain)  G^{eik,\,(l-1)}_{{\bf a}c}(k;\,\eta,s) \big[ \Phi^{eik}_{cd}(k;s,s') G^{eik}_{bd}(k;\,\eta,s')\big]^{(n-l)}\nonumber\\
&&\qquad\qquad\quad+    \big[ G^{eik}_{ac}(k;\,\eta,s) \Phi^{eik}_{cd}(k;s,s') \big]^{(n-l)} G^{eik,\,(l-1)}_{{\bf b} d}(k;\,\eta,s') H^{eik}_{\bf b}(k;\,\eta,\etain)  \Big\}\nonumber\\
&&+ \int \, ds'\;\Big[\Phi^{eik, L}_{ac}(k;\eta,s') G^{eik}_{bc}(k;\,\eta,s')+ G^{eik}_{ac}(k;\,\eta,s')\Phi^{eik, R}_{cb}(k;s',\eta) \Big]^{(n)}\,,
\label{Teik}
\eeqra
where $H^{eik}_a(k;\,\eta,\etain) = \int_{\etain}^\eta ds' \Sigma^{eik, \,(1)}_{ae}(k; \eta, s')u_e = - u_a\,k^2 \sigma_v^2\, e^\eta(e^\eta-e^{\etain})$.

The first two lines at the RHS of \re{Teik}, summed over $n$ and combined with the second line of ~\re{Texact}, give
\beq
 H^{eik}_{\bf a}(k;\,\eta,\etain)  \,P^{eik}_{{\bf a} b}(k; \eta)  + H^{eik}_{\bf b}(k;\,\eta,\etain)  \,P^{eik}_{a{\bf b}}(k; \eta)\,,
\eeq
where we have used the fact that  $P^{eik}_{ab}$ has the structure given in eq.~\re{fullP}, with the full $G_{ab}$ and $\Phi_{ab}$ replaced by their eikonal limits.

The functions  $\Phi^{eik, L}_{ab}$ and  $\Phi^{eik, R}_{ab}$ are obtained from the full $\Phi^{eik}_{ab}$ by taking into account only those contributions having a tree level, {\it i.e.} not-renormalized, vertex at the end corresponding to the index ``a" and ``b", respectively.  
The first term in the last line of eq.~\re{Teik}  can be schematically represented as in the LHS of fig.~\ref{PhiGeik}. There, the horizontal thick lines represent eikonal propagators as defined in eq.~\re{resCS}, carrying the hard momentum $k$,  whereas the thin lines represent soft propagators, which can be well approximated by the linear $g_{ab}$.

\begin{figure}
\centerline{\includegraphics[width = 16cm,keepaspectratio=true]{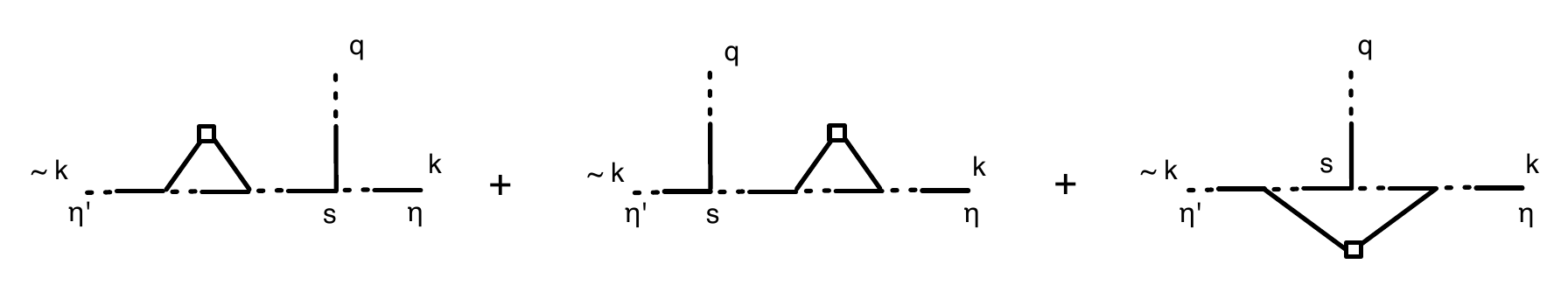}}
\caption{The one-loop correction to eq.~\re{vtreegg} }
\label{1lvergg}
\end{figure}

The $m$ lines on the left have to be joined to the $m$ ones on the right in all possible ways, by means of $m$ soft PS's. Notice that the number $m$ in the figure  does not correspond  to the order in standard PT, since the thick lines and the thick dots already include infinite orders in standard PT. In particular, the thick dots represent fully renormalized vertices in the eikonal limit. These vertices are obtained by correcting the tree level expression
\beqra
&& g_{ad}(\eta,s) e^s \gamma_{dce}(\bk,-\bq,\bq-\bk) u_c\, g_{eb}(s,\etap)\nonumber\\
&&\qquad \to g_{ab}(\eta,\etap)\, e^s\frac{1}{2} \frac{\bk \cdot \bq}{q^2 }\,\Theta(\eta-s)\Theta(s-\etap) \qquad\qquad (\mathrm{for\; k\gg q})\,,
\label{vtreegg}
\eeqra 
by including all possible soft PS insertion on the hard propagator lines. At 1-loop, the expression above is corrected by the three diagrams in fig.~\ref{1lvergg}, which, again in the eikonal limit, sum up to,
\beq
- k^2\sigma_v^2 \frac{(e^\eta-e^{\etap})^2}{2} \, g_{ab}(\eta,\etap)\,\times\, e^s\frac{1}{2} \frac{\bk \cdot \bq}{q^2 }\Theta(\eta-s)\Theta(s-\etap) \,,
\eeq   
where we recognize, at the first factor, the 1-loop contribution to $G^{eik}_{ab}(k;\eta,\etap)$.
Considering all loop orders, and summing up, one finds
\beqra
&&\int ds_1\,ds_2 \, G^{eik}_{ad}(k;\eta,s_1) \Gamma^{eik}_{dce}(\bk,-\bq,\bq-\bk;s_1,s,s_2) u_c\,G^{eik}_{eb}(|\bq-\bk|;s_2,\etap) \nonumber\\
&&\qquad \to  G^{eik}_{ab}(k;\eta,\etap) \,e^s\frac{1}{2} \frac{\bk \cdot \bq}{q^2 }\,\Theta(\eta-s)\Theta(s-\etap)
\qquad(\mathrm{for\; k\gg q})\,,
\label{gammaren}
\eeqra 
where $ \Gamma^{eik}_{dce}$ is the fully renormalized vertex in the eikonal limit, represented as a thick dot in fig.~\ref{PhiGeik}. Notice that the leftmost (rightmost, if $\Phi^R_{ab}$ is considered) vertex is not thick, {\it i.e. } renormalized, but it is given by the tree level $\gamma_{abc}$ of eq.~\re{vertice}.

Therefore, in the eikonal limit, the time integrations on the insertion points of the soft legs via a full vertex, {\it i.e.} those corresponding to the thick dots in fig.~\ref{PhiGeik}, factorize from the time integrations of the loops correcting $g_{ab}(\eta,\etap)$ to $G^{eik}_{ab}(k;\eta,\etap)$, and can be performed independently, provided the causal ordering enforced by the theta-functions in eq.~\re{gammaren} is respected.

The contributions obtained by taking into account the $m!$ possible pairings between the soft lines by  the $m$ soft PS's in fig.~\ref{PhiGeik}, and then by integrating the $m-1$ full vertices on the left and the $m$ ones on the right from $\etain$ to $\eta$,  give
\beq
G^{eik}_{ac}(k;\eta,\etain) G^{eik}_{bd}(k;\eta,\etain) P^0(k) \,u_c u_d \frac{\big[k^2 \sigma_v^2 (e^\eta-e^{\etain})^2 \big]^m} {m!\,(m-1)!} m!\,,
\eeq
which, summed on $m$ from 1 to $\infty$ gives the simple result
\beqra
&& \int\, ds\;\big[\Phi^{eik,\,L}_{ad}(k;\eta,s)  G^{eik}_{bd}(k;\,\eta,s)+G^{eik}_{ad}(k;\,\eta,s)\Phi^{eik,\,R}_{db}(k;s,\eta) \big] \nonumber\\
&&= 2 P^0(k) u_a u_b k^2 \sigma_v^2  e^{2 \eta}\,,
\label{eik3}
\eeqra
where we have taken the $\etain\to -\infty$ limit.

In the eikonal limit, the evolution equation \re{Texact} can then be rewritten as
\beqra
&&\partial_\eta \,P^{eik}_{ab}(k; \eta) =-\Omega_{ac} \,P^{eik}_{cb}(k; \eta)  -\Omega_{bc} \,P^{eik}_{ac}(k; \eta)\nonumber \\
&&\qquad\qquad\qquad  + H^{eik}_{\bf a}(k;\,\eta,\etain)  \,P^{eik}_{{\bf a} b}(k; \eta)  + H^{eik}_{\bf b}(k;\,\eta,\etain)  \,P^{eik}_{a{\bf b}}(k; \eta) \nonumber \\
&&\qquad \quad+\int_{\etain}^\eta ds\;\big[\Phi^{eik,\,L}_{ad}(k;\eta,s)  G^{eik}_{bd}(k;\,\eta,s)+G^{eik}_{ad}(k;\,\eta,s)\Phi^{eik,\,R}_{db}(k;s,\eta) \big] \nonumber\\
&&\quad\qquad\qquad= -\Omega_{ac} \,P^{eik}_{cb}(k; \eta)  -\Omega_{bc} \,P^{eik}_{ac}(k; \eta)\nonumber \\
&& \qquad\qquad\qquad -2 \,k^2\, \sigma_v^2 \,e^{2\eta}\,P^{eik}_{ ab}(k; \eta)  \nonumber\\
&&\qquad\qquad\qquad+ 2 P^0(k) u_a u_b k^2 \sigma_v^2  e^{2 \eta}\,,
\label{Tstartv}
\eeqra
where we have consistently taken $H_{ a}\to H^{eik}_a$ also at the second line of \re{Texact}. Notice the formal similarity between eq.~\re{Tstartv} and the equations we derived for the low-$k$ regime, eqs.~\re{TR1}, \re{TR2}, suggesting that, as for the propagator \cite{Anselmi:2010fs}, an equation of the same structure could be used to interpolate between the two extremal regimes.

The equation above has the attractor solution 
\beq
P^{eik}_{ab}(k; \eta) = P^0(k) u_a u_b\,,
\label{peik}
\eeq
as can be verified directly, by using $\Omega_{ac} u_c=0$ (see eqs. ~\re{bigomega} and \re{ic}).
This result can be generalized to the two-point correlator at non-equal times $\eta$ and $\etap$. The relevant diagrams to compute this quantity are obtained  by attaching  an eikonal propagator (connecting the time $\eta$ to $\etap$) to the left end of the LHS of fig.~\ref{PhiGeik}  and by replacing the leftmost tree level vertex  by a thick one. Proceeding analogously to what we did above, one gets
\beq
P^{eik}_{ab}(k; \eta,\etap) = P^0(k) u_a u_b \exp\left[-k^2 \sigma_v^2 \frac{(e^\eta-e^{\etap})^2}{2}\right]\,,
\label{Pnle}
\eeq
which reduces to \re{peik} for $\eta=\etap$.
 
The equality between the fully nonlinear  PS, in the eikonal limit,  and the linear one\footnote{We are considering a single fluid with initial conditions in the adiabatic growing mode, for the case of isocurvature modes, see \cite{Bernardeau:2011vy, Bernardeau:2012aq}.}, can be understood physically as due to the fact that all the soft modes, in the extreme $q\ll k$ limit, are seen, at the short scale set by $k$, as zero modes. Since, by Galileian invariance,  equal-time correlators should be independent on zero modes \cite{Scoccimarro:1995if}, all the nonlinear corrections cancel out in this limit (see also  \cite{Bernardeau:2012aq}).

In order to outline the interpolation procedure between the large and the small $k$ regimes, it is useful to notice that the contributions leading to eq.~\re{eik3}, once summed over $m$, can also be represented by the two diagrams on the RHS of fig.~\ref{PhiGeik}, where the black PS now is the fully renormalized two-point correlator in the eikonal limit, eq.~\re{Pnle}, and all the vertices besides the leftmost one are fully renormalized (in the eikonal limit). The first diagram comes from all pairings in which the vertex in $\eta$ is paired to that in $s'$, the second one from all the other ones.

\subsection{The intermediate $k$ regime}
\label{intk}
The eikonal limit provides a conceptually interesting benchmark, but it does not adequately describe the large $k$ limit. One one hand, in practical situations such as $\Lambda$CDM, the separation of scales is not so abrupt  as to motivate an effectively $q\to 0$ limit for the soft modes when the hard one is $k=O(1) \;{\mathrm{h\,Mpc}^{-1}}$. On the other hand, considering arbitrarily large momenta is meaningless because of multi streaming effects, which are totally neglected in the PT framework, but are certainly relevant for $k \agt 1 \;{\mathrm{h\,Mpc}^{-1}}$ and higher \cite{Pueblas:2008uv, Valageas:2010rx, Pietroni:2011iz}.

In order to recover the proper momentum dependence for the nonlinear propagator at lower k, in \cite{Anselmi:2010fs} we used the 1-loop expression for $H_{ a}$ in the evolution equation \re{dgtot}. The equation so obtained provides an interpolation between the large $k$ limit, in which $H^{eik}_{ a}$ generates the all-order resummed propagator $G^{eik}_{ab}$,  and 1-loop standard PT at small $k$. 

In this paper we will follow an analogous path to get an interpolation procedure for the nonlinear PS. First, we will assume that the evolution equation has the same structure as eqs.~\re{TR1}, \re{TR2}, and \re{Tstartv}, for any values of $k$, and use appropriate values for the functions $H_{ a}$, $G_{ab}$, and $\Phi^{L(R)}_{ab}$,  appearing in those equations.
As for $H_{ a}$ we will use the complete 1-loop expression, as we did in \cite{Anselmi:2010fs}. In principle $G_{ab}$, appearing at the third line of the evolution equation, should be taken equal to $\bar{G}_{ab}$ introduced in eqs.~\re{G1loop}, \re{resCS}, however we checked numerically that using the more numerically convenient $G_{ab}^{eik}$ gives undistinguishable results. Then, it remains to be discussed the interpolating expression for  the mode coupling function $\Phi^{L(R)}_{ab}$, between the low-$k$ value, $\Phi^{(1)}_{ab}$ and the eikonal limit. Before doing that, as a warm-up, we consider again the evolution equation for the propagator, eq.~\re{dgtot}. 

\begin{figure}
\centerline{\includegraphics[width = 10cm,keepaspectratio=true]{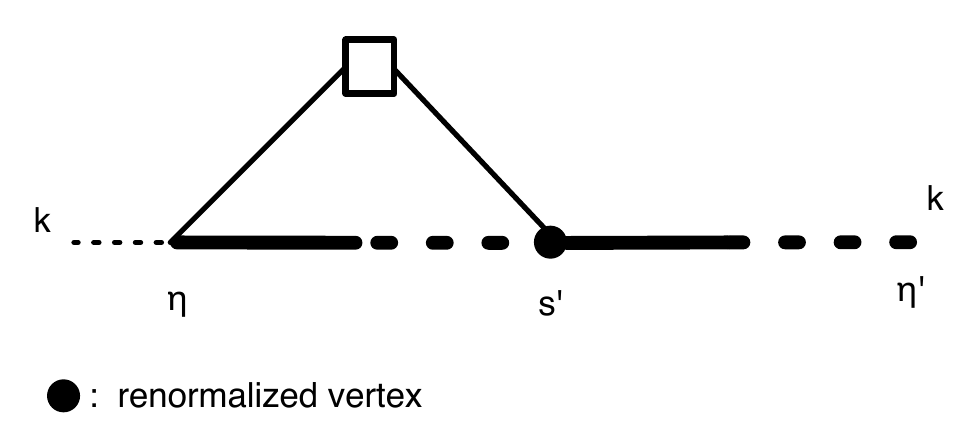}}
\caption{The diagram for $\Sigma^{eik, L}_{ac}(k;\eta,s')  G_{cb}^{eik}(k;c',\eta')$.}
\label{geik}
\end{figure}

\begin{figure}
\centerline{\includegraphics[width = 8cm,keepaspectratio=true]{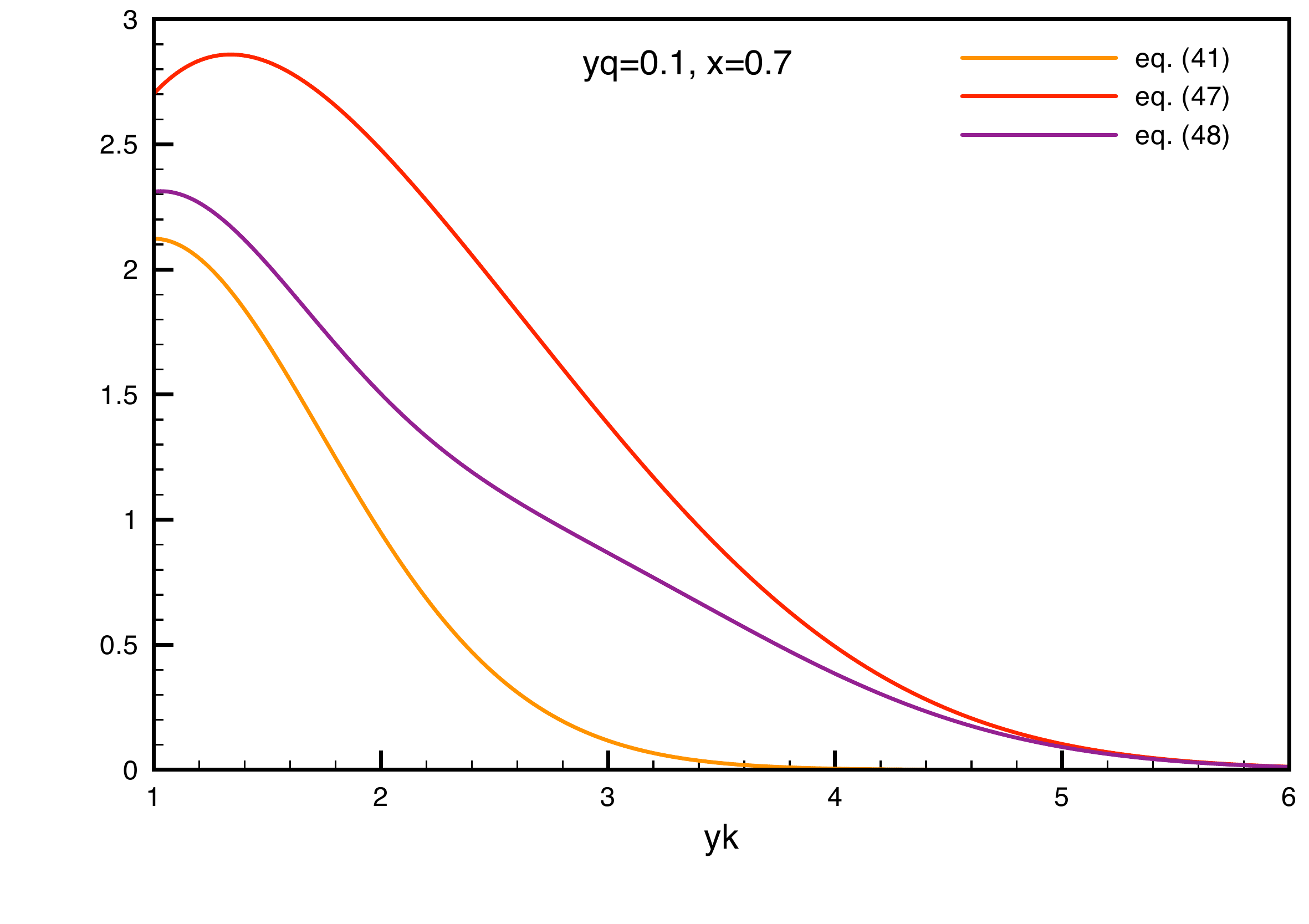}\includegraphics[width = 8cm,keepaspectratio=true]{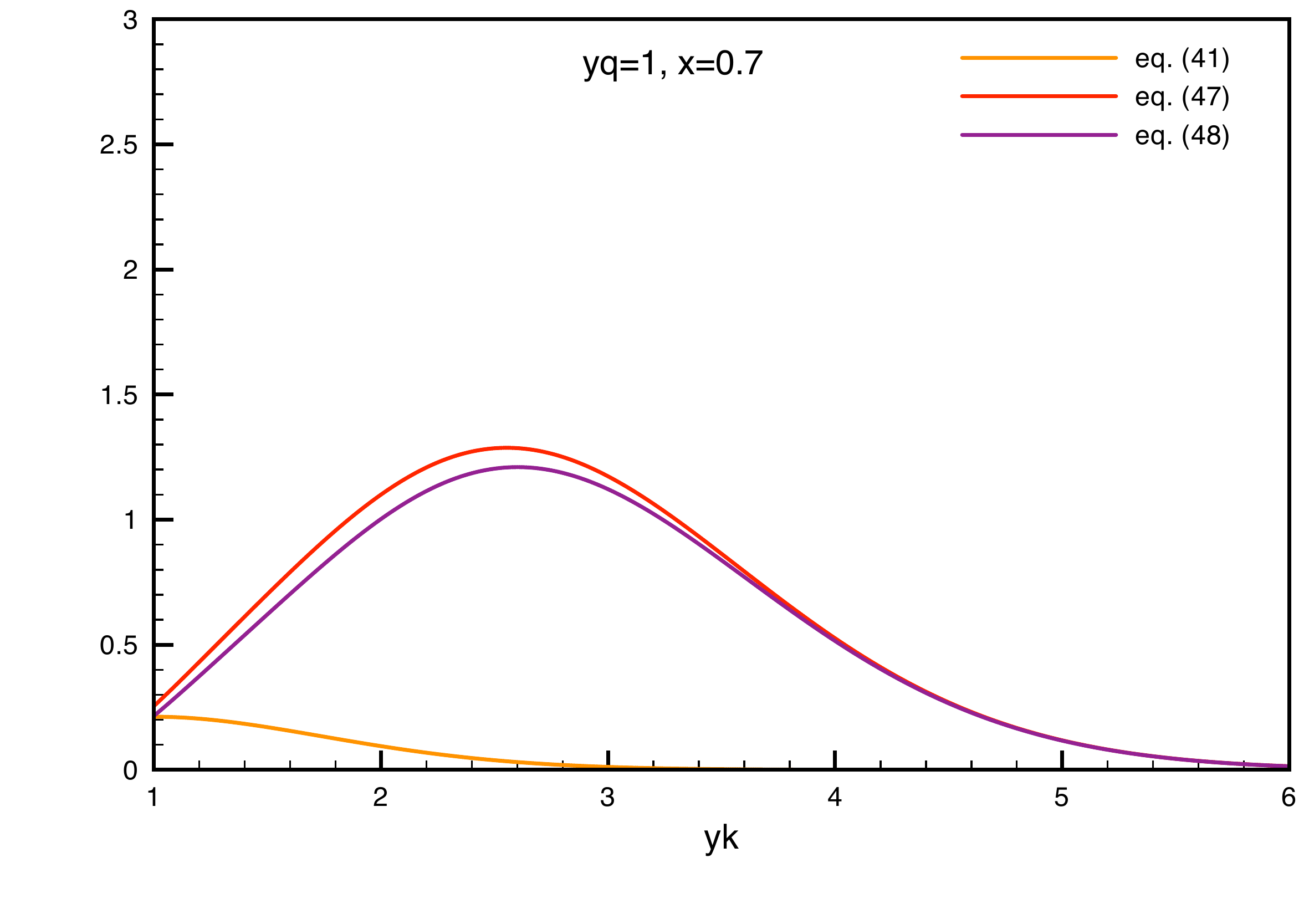}}
\centerline{\includegraphics[width = 8cm,keepaspectratio=true]{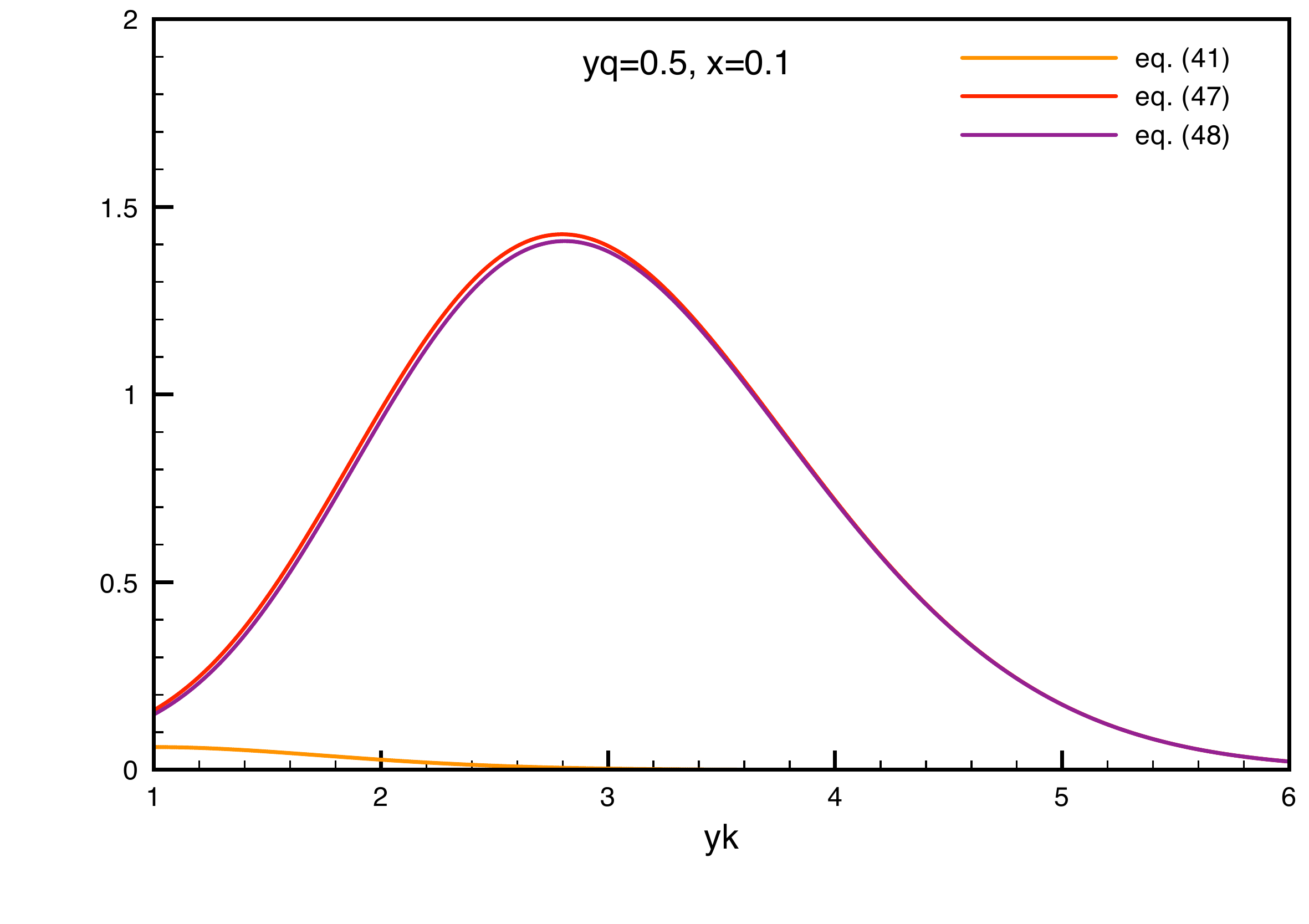}\includegraphics[width = 8cm,keepaspectratio=true]{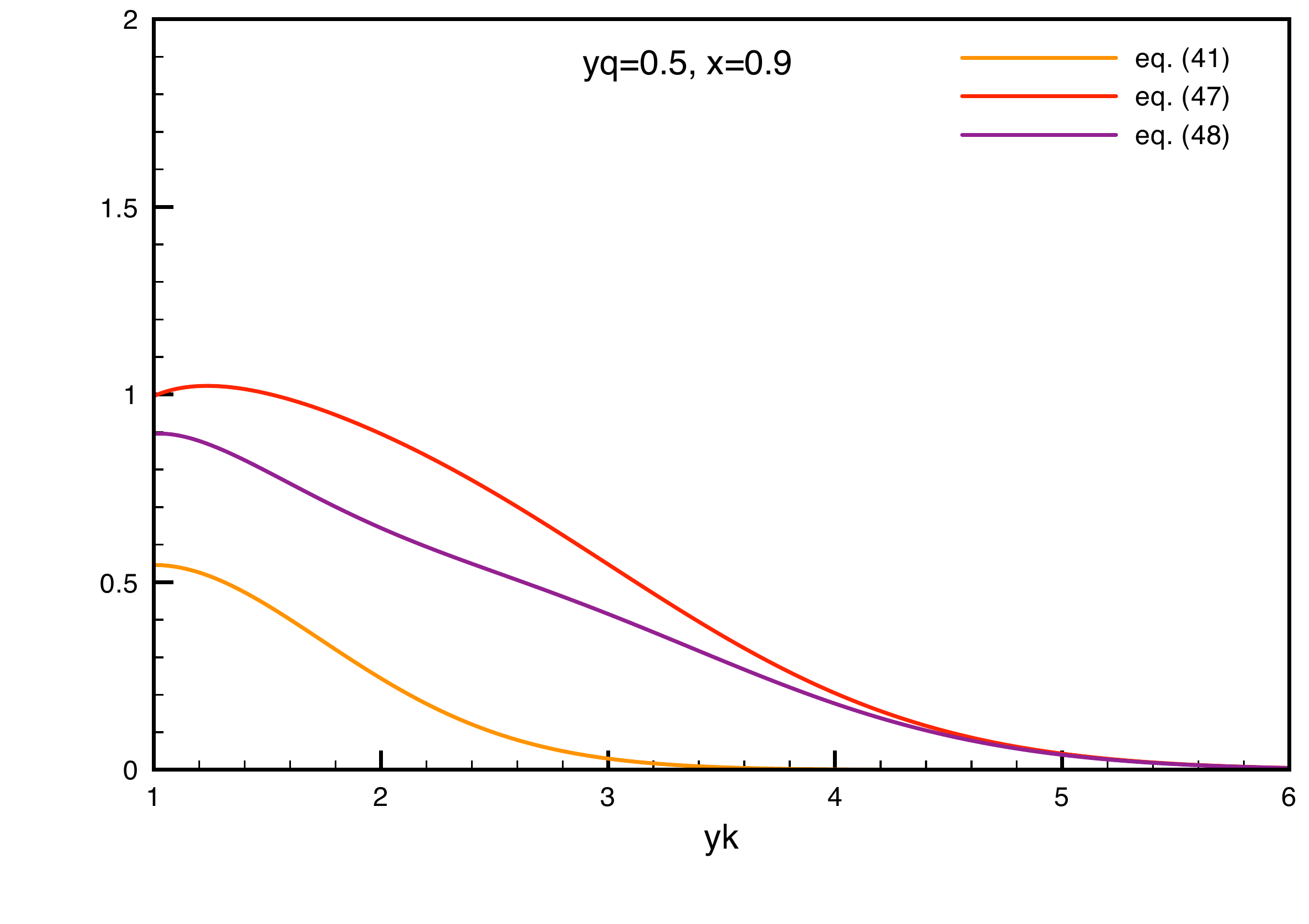}}
\caption{The improved vertex interaction of eq.~\re{impv} (purple line), the full eikonal expression of eq.~\re{gammaren} (orange line), and the tree-level vertex interaction of eq.~\re{gtree} (red line), for different values of $y_q$ and $x=\bk\cdot\bq/k q$. Notice that, at $z=0$, $y_q=0.1,0.5,1$ correspond to $q=0.016,0.08,0.16\;\mathrm{h\,Mpc^{-1}}$, respectively.}
\label{vertici}
\end{figure}

In the evolution equation for the propagator, a time-integrated product between $\Sigma_{ac} $ and $G_{cb}$, see eq.~\re{deltaG}, appears, which in the eikonal limit takes the factorized expression in eq.~\re{TRGL}. This quantity can be represented by just the 1-loop diagram of fig.~\ref{geik}, where the hard propagators and the rightmost vertex are the fully renormalized, eikonal ones. Notice that, being the vertex at $s'$ a renormalized one, the two hard propagators combine in a single one, and the integration over $s'$ factorizes (see eq.~\re{gammaren}), leading to eq.~\re{TRGL} with $H_a=H_a^{eik}$. Now, to get the correct $k\to 0$ limit from this expression one has to replace the fully renormalized vertex with a tree level one, obtaining in this case eq.~\re{TRGL} plus subleading  $O(k^4 \sigma_v^4)$ corrections. This is the same result that one would obtain in eRPT at 1-loop, see \ref{eRPT}. Therefore, the interpolation between large and small $k$ can be entirely loaded on the shoulders of vertex renormalization. Indeed, at the scales of interest, vertex renormalization is not fully at work as to give the eikonal expression in eq.~\re{gammaren}. This is shown in fig.~\ref{vertici}, where we plot eq.~\re{gammaren} (orange line), the corresponding quantity in which the thick vertex has been replaced by a tree-level vertex (red line), 
\beq
G^{eik}_{ad}(k;\eta,s) e^s \gamma_{dce}(\bk,-\bq,\bq-\bk) u_c\, G^{eik}_{eb}(k;s,\etap)\,,\label{gtree}
\eeq
and the improved interaction term 
\beqra
{\mathrm{eq.\;\re{gammaren}}} \,&-& \,G^{eik}_{ad}(k;\eta,s) e^s \frac{1}{2}\frac{\bk\cdot\bq}{q^2}\, G^{eik}_{db}(k;s,\etap)\nonumber\\
&+& G^{eik}_{ad}(k;\eta,s) e^s \gamma_{dce}(\bk,-\bq,\bq-\bk) u_c\, G^{eik}_{eb}(k;s,\etap)\,,
\label{impv}
\eeqra
(purple line) obtained by replacing the tree-level vertex contribution to \re{gammaren} in the eikonal limit with the tree level vertex in its complete form $ \gamma_{dce}(\bk,-\bq,\bq-\bk) $. Of course eq.~\re{impv} goes to eq.~\re{gammaren} in the eikonal limit, but the question here is to understand how far from this limit it is at the scales of interest in the paper. 

All the three expressions are computed setting the index $a=1$, and have been  contracted with $u_b$ and integrated in $s$ from $\etap \to -\infty$ to $\eta$. Moreover, we have divided by $\exp(\eta)$. The resulting quantities depend on $y_k=k \sigma_v e^\eta$, $y_q=q \sigma_v e^\eta$, and $x= \bk\cdot\bq/k q$. The plots in fig.~\ref{vertici} clearly show that the contribution from vertex renormalization (in the eikonal limit) is subdominant for $y_k$ up to 6, corresponding, at $z=0$ to $1\;\mathrm{h\,Mpc^{-1}}$. Only for very small values of the soft momentum, the eikonal limit for the vertex interaction is approached.

We are therefore motivated to extend the procedure already discussed above from the $\Sigma_{ac}$ to the $\Phi^{L (R)}_{ab}$, namely,  neglecting vertex renormalization. This is further justified by the fact that the soft momentum contributing to $\Phi_{ab}$ is typically larger than those dominating the $\Sigma_{ab}$, due to the presence of a $P^0(q)$ in the loop. The prescription of using tree level vertices is also motivated by eRPT, as discussed in \ref{eRPT}. 

Looking at the RHS of fig.~\ref{PhiGeik}, we have two contributions. If we replace the full vertex in the first diagram with a tree-level one we get the 1-loop expression in eRPT for $\Phi_{ab}$ of eq.~\re{phi1erpt}, which in the $k\gg q$ limit can be approximated as
\beq
\Phi^{eRPT, (1)}_{ab}(k;s,s')\simeq \Phi^{(1)}_{ab}(k;s,s')  \exp\left[-k^2 \sigma_v^2 \frac{(e^s-e^{s'})^2}{2}\right]\,,
\label{phie1}
\eeq
where $\Phi^{ (1)}_{ab}(k;s,s')$ is computed in 1-loop standard PT. 
As for the second contribution we also replace the rightmost vertex with a tree-level one. In order to avoid 2-loop integrations, we further simplify this contribution by taking the tree level vertices in their limiting form, eq.~\re{vtreegg}, and take $P^0(|\bk-\bq|)\simeq P^0(k)$ inside the loops. We get
\beq
\left(k^2 \sigma_v^2 \,e^{s+s'} \right)^2 P^0(k) u_a u_b \,\exp\left[-k^2 \sigma_v^2 \frac{(e^s-e^{s'})^2}{2}\right]\,.
\label{phie2}
\eeq
 Since both contributions, eqs.~\re{phie1} and \re{phie2}, have tree level vertices at both ends, there is no difference between $\Phi^L_{ab}$ and $\Phi^R_{ab}$ at this level. The latter will therefore both be approximated by the sum of eqs.~\re{phie1} and \re{phie2}, namely,
\beq
\tilde{ \Phi}_{ab}(k;s,s') \equiv e^{-\frac{k^2 \sigma_v^2}{2} (e^{s}-e^{s'})^2}\,\left[\Phi_{ab}^\mathrm {(1)}(k;s,s^\prime)+\left(k^2 \sigma_v^2 \,e^{s+s'} \right)^2 P^0(k) u_a u_b \right]
 \,.
\label{largephitnew}
\eeq
At small $k$, the above expression explicitly goes to $ \Phi_{ab}^\mathrm {(1)}+O(k^4)$ .
The second term inside parentheses comes from the second diagram at the RHS of fig.~\re{PhiGeik} and it is crucial in order to recover the eikonal limit at large $k$, once vertex renormalization for the rightmost vertex is restored, as in fig.~\ref{PhiGeik}. However, it should be switched-off for small momentum values, because in that range it becomes degenerate with 2-loop diagrams not included in this resummation, which would give other $O(k^4)$ contributions. In order to avoid taking into account these new contributions, we suppress the second term in eq.~\re{largephitnew}  at small $k$ multiplying it by a filter function of the form
\beq
F(k)= \frac{\left(k/\bar k\right)^4}{1+\left(k/\bar k\right)^4}\,,
\label{filter}
\eeq
see also eq.~\re{ufilter}, where the power $4$ ensures a rapid enough switch-off of the filter at low $k$. 

As we will see in next section, see fig.~\ref{PS_Large}, the filter becomes irrelevant at large $k$ and for redshifts $\agt 1$. Therefore we fix $\bar{k}$ at $z=0$ (where the filter effect is maximal) by taking it equal to the scale $k$ at which the two terms of eq.~\re{largephitnew} are equal.  Choosing a smaller value would suppress the 1-loop contribution to eq.~\re{largephitnew} which is shown to successfully reproduce the BAO scales by our discussion of the small $k$ range (see approx. ``R2" introduced in sect.~\ref{smallk}). On the other hand, choosing a larger $\bar k$, would delay too much (in $k$) the effect of the large $k$ contribution. This procedure gives the value  $\bar{k}=0.2\; \mathrm{h \,Mpc^{-1}}$, that we will use explicitly in our numerical computation. In the next section we will also show results in which no filter function is used in eq.~\re{largephitnew}, see fig.~\ref{PS_Large}, and in \ref{AB} we will show results obtained by varying the value of the filter w.r.t. the optimal value.

In summary, in the following we will solve the evolution equation 
\beqra
\partial_\eta \,P_{ab}(k; \eta) &=& -\Omega_{ac} \,P_{cb}(k; \eta)  -\Omega_{bc} \,P_{ac}(k; \eta)\nonumber \\
&& + H_{{\bf a}}(k;\, \eta,\etain)\,P_{{\bf a}b}(k; \eta) +H_{{\bf b}}(k;\, \eta,\etain)\,P_{a{\bf b}}(k; \eta) \nonumber\\
&&+ \int \, ds\;\big[\tilde{\Phi}_{ad}(k;\eta,s) G^{eik}_{bd}(k;\eta,s)+ G^{eik}_{ad}(k;\eta,s)\tilde{\Phi}_{db}(k;s,\eta) \big]\,, \nonumber\\
\label{TReik}
\eeqra
where $H_{{a}}(k;\, \eta,\etain)$ is given in eq.~\re{TRGL} and $\tilde{\Phi}_{ad}(k;\eta,s)$ is given in eq.\re{largephitnew} with the filter function \re{filter} inserted as described above. Explicit expressions for these functions are given in \ref{AB}.

We reiterate that the above equation interpolates between the small $k$ and the large $k$ limits discussed in the previous sections.
For small $k$, we get the approximation R2 discussed in Sect.~\ref{smallk}\footnote{Actually, to get exactly R2 in the small $k$ limit, we should use $\bar G$ instead of $G^{elk}$ at the last line of eq.~\re{TReik}, however the two choices give sub percent differences, while the use of $G^{elk}$ is much more convenient numerically.}. 

The large $k$, eikonal limit, is recovered by replacing back the full vertex $\Gamma^{eik}_{abc}$ for the tree one $\gamma_{abc}$ at the rightmost (the leftmost one, if $\Phi^{R}_{ab}$ is considered) vertices in fig.~\ref{PhiGeik}. This operation turns the last line of eq.~\re{TReik} into the eikonal limit of eq.~\re{Tstartv}. However, since, as we discussed above, the eikonal limit is not realized in practice, even at very large $k$, we will not introduce an explicitly interpolating procedure between these two expressions, but will use eq.~\re{TReik} for all values of $k$, using the comparison with N-body simulations to assess the range of validity of the approximation. An alternative, more systematic but also more computing time consuming approach, would be to consider the improved vertex in \re{impv} and/or  higher orders in eRPT. We leave the investigation of this way for future work.

It is therefore instructive to study the very large $k$ limit of eq.~\re{TReik}. Using the large $k$ limit for the first term of eq.~\re{largephitnew}, 
\beq
\Phi^{eRPT, (1)}_{ab}(k;s,s') \to \left(k^2 \sigma_v^2 \,e^{s+s'} \right) P(k) u_a u_b \,\exp\left[-k^2 \sigma_v^2 \frac{(e^s-e^{s'})^2}{2}\right]\,,
\label{phie}
\eeq
the last line of eq.~\re{TReik} can be integrated analytically in $s$, to give
\beq
 u_a u_b P^0(k) \;\big[ y^2 (e^{-y^2}-1)+\sqrt{\pi}\; y (1+y^2) \mathrm{Erf}(y)\big],\qquad\mathrm{where} \;y\equiv e^\eta k\sigma_v\,,
\label{phig}
\eeq
and ``Erf" is the error function, $\mathrm{Erf}(y)\equiv2\pi^{-1/2}\,\int_0^y dt \,e^{-t^2}$.
At large $y$, the above expression goes as $P^0(k) \sqrt{\pi} y^3(1-1/y\sqrt{\pi}+1/y^2)$. Considering that $H_a$ and $H_b$, in the same regime, go as $- y^2$, we realize that in the large-$y$ limit  the differential equation in (\ref{TReik}) has an attractor solution, given by the simple formula
\beq
P_{ab}(k;\eta) \to \frac{\sqrt{\pi}}{2} y \left(1-\frac{1}{y\sqrt{\pi}}+\frac{1}{y^2}\right) \, u_a u_bP^0(k)\,,\qquad\qquad \mathrm{for \;large }\, y\,,
\label{approxasy}
\eeq
which exhibits the pure $y$ dependence for the nonlinear to linear PS ratio we found for the N-body simulations, see fig.~\ref{selfsim}. 

In sect.~\ref{results}, we will present results for the solutions of eq.~\re{TReik} as well as for the analytic approximation \re{approxasy}.

\section{Numerical results: comparison with N-body simulations}
\label{results}

\begin{figure}
\centerline{\includegraphics[width = 17cm,keepaspectratio=true]{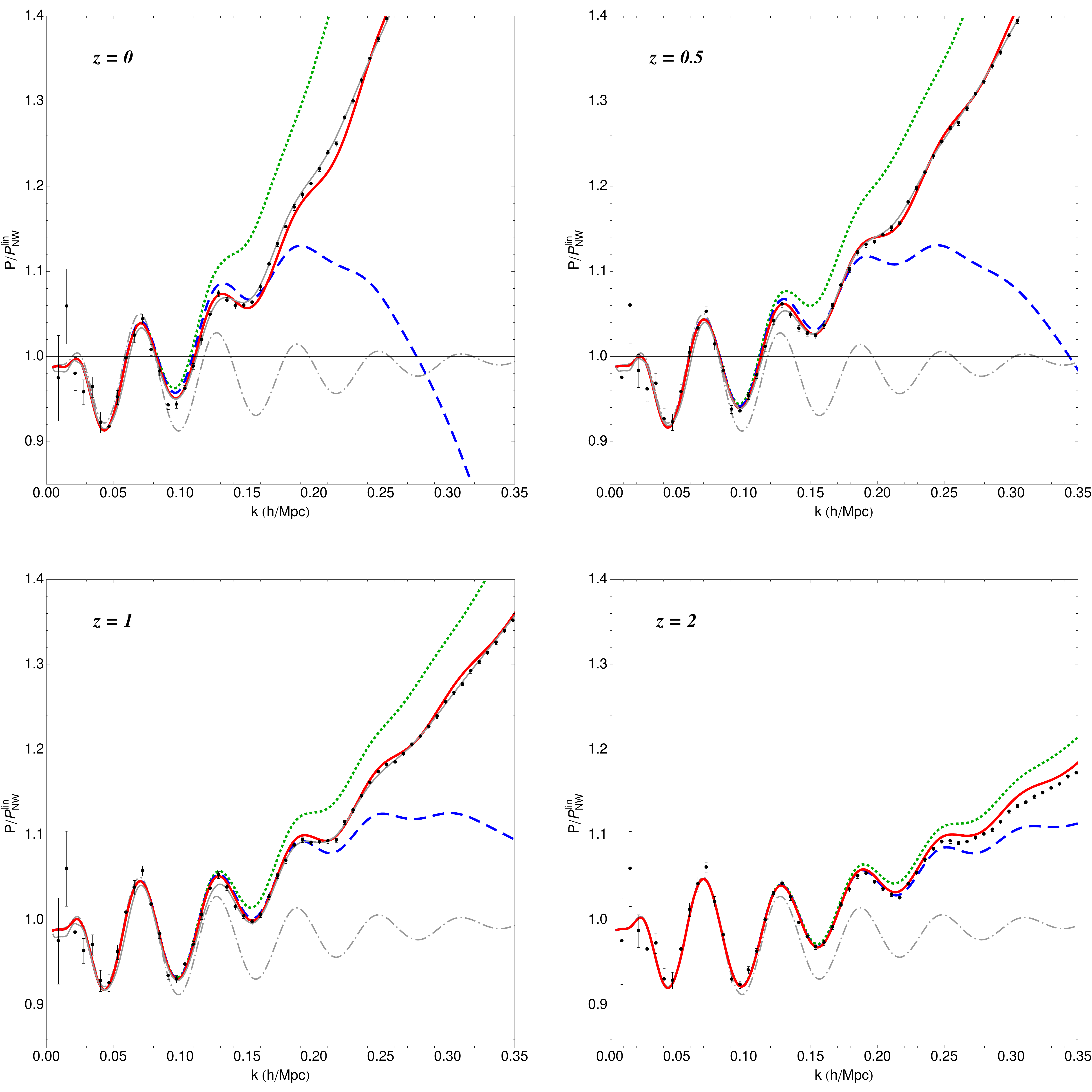}}
\caption{PS's normalized to the no-wiggle PS of \cite{eisensteinhu} at different redshifts. The color code is the following: the grey dash-dotted line is linear PT, the green dotted line is 1-loop PT, the blue dashed line corresponds to the solution of the evolution equation in the small $k$ approximation ``R2" of eq.~\re{TR2}, the red solid line is the solution of the evolution equation \re{TReik}, and dots with error-bars are the N-body results of  Sato and Matsubara \cite{Sato:2011qr}. Also shown (thin grey line) is the output of the Coyote interpolator of refs.~\cite{Heitmann:2008eq, Heitmann:2009cu, Lawrence:2009uk}.}
\label{PS_BAO}
\end{figure}

\begin{figure}
\centerline{\includegraphics[width = 17cm,keepaspectratio=true]{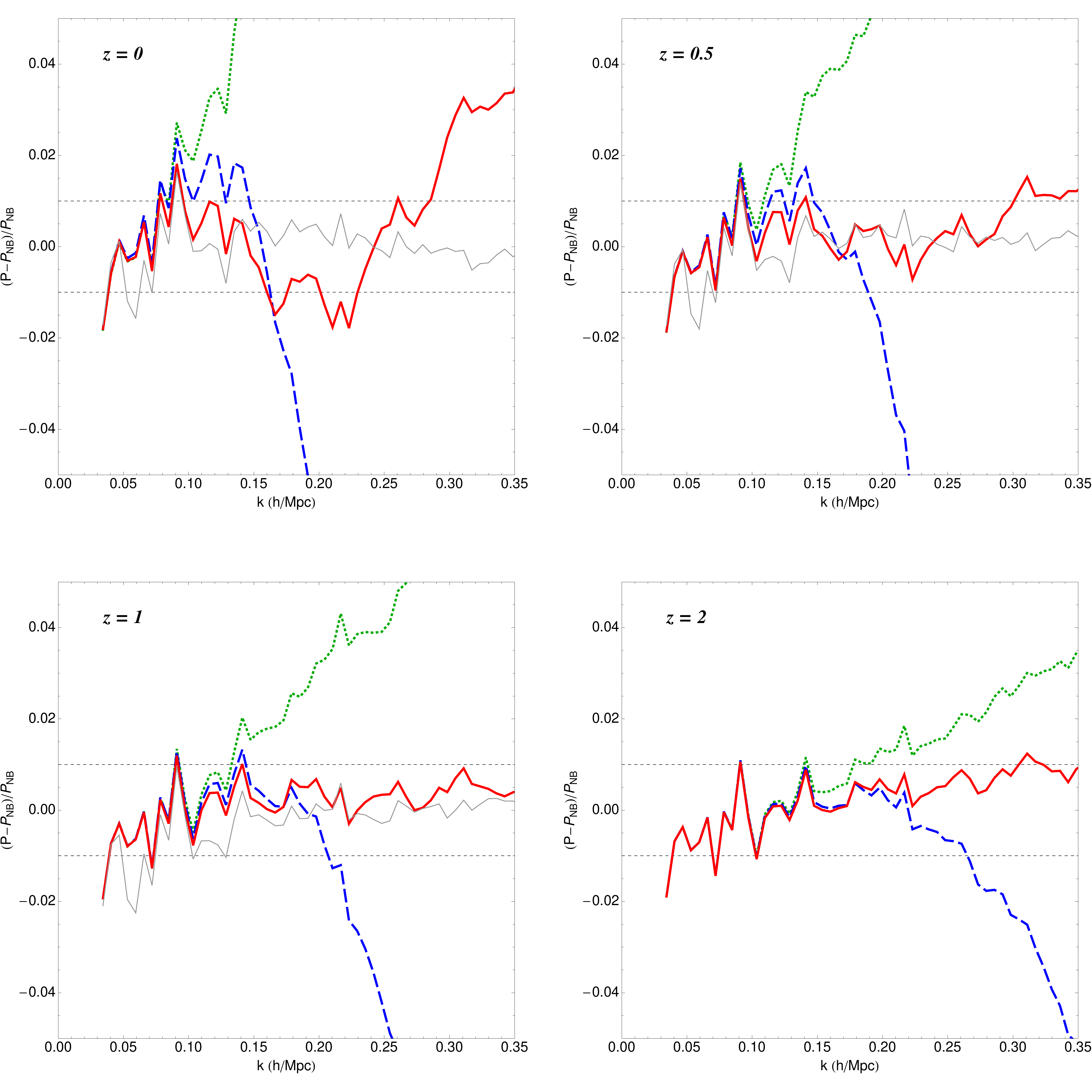}}
\caption{The relative difference between various approximations discussed in the text and the N-body simulations of Sato and Matsubara \cite{Sato:2011qr}. Same color-code as in fig.~\ref{PS_BAO}. The thin grey line is the comparison with the output of the Coyote interpolator. }
\label{PS_BAO_diff}
\end{figure}

The aim of this section is to test numerically the goodness of our approach. We compared the nonlinear PS predictions against the data coming from high accuracy N-body simulations designed to predict the nonlinear PS at the percent accuracy around the BAO range of scales. The initial linear PS for the simulations we considered was obtained from the CAMB public code \cite{CAMB} and any primordial non--Gaussianity was neglected. 
Accordingly, we solved our evolution equations taking the initial PS from CAMB at  $z_{in}=100$, where the gravitational clustering is fully linear on the scales of interest.

We provide plots of the comparison between our approach and the N-body simulations presented by Sato and Matsubara in \cite{Sato:2011qr}. They considered a $\Lambda$CDM cosmology with the following  parameters: $\Omega_{m}=0.265$, $\Omega_{b}h^2=0.0226$, $h=0.71$, $n=0.963$ and $\sigma_{8}=0.8$. In order to check the possible dependence of our results on the cosmology or on the N-body simulations, we also compared our results with the independent set of simulations produced by Carlson, White and Padmanabhan in \cite{Carlson:2009it} for a different $\Lambda$CDM cosmology, and with the results from the cosmic Coyote emulator \cite{Heitmann:2008eq, Heitmann:2009cu, Lawrence:2009uk}. The latter is an interpolator built from a suite of many different N-body simulations which provides nonlinear PS's for different cosmological parameters chosen inside a certain range and for redshifts $z\le 1$. In all these cases our comparison tests worked at the same quality level.

In fig.~\ref{PS_BAO} we plot the nonlinear PS computed in different approximations, divided by the smooth linear PS given in  \cite{eisensteinhu}.  The blue dashed line corresponds to the small $k$ approximation of eq.~\re{TR2}, while the red solid line is obtained by using eq.~\re{TReik}. The results obtained in linear PT (grey dash-dotted) and 1-loop PT (been dotted) are also shown. Relative differences with respect to the N-body results are given in fig.~\ref{PS_BAO_diff} with the same color-code. 

We notice that our evolution equation in the small $k$ limit of eq.~\re{TR2}, {\it i.e.} with no-resummation of the contributions to $\Phi_{ab}$ beyond 1-loop order,  is able to reproduce the nonlinear PS at the percent level in the BAO range. There are a couple of outlier points in the linear region $k<0.1\;\mathrm{h/Mpc}$ which can be ascribed to a fluctuation in the N-body simulations of \cite{Sato:2011qr} from a comparison with the Coyote emulator (thin grey line in fig.~\ref{PS_BAO},  \ref{PS_BAO_diff}) and with the 1-loop approximations, which works well in this region.
Concerning the resummed and interpolated solution, {\it i.e.} the red line, we see that it performs at the $1 \%$ level in the BAO region at any redshift, including $z=0$, where, due to the larger amount of nonlinearity, the last peak is at about $0.15\;\mathrm{h/Mpc}$. At redshifts $z\ge 0.5$ it performs as the Coyote interpolator up to $k\simeq 0.35 \;\mathrm{h/Mpc}$, namely, still in agreement with the N-body results of  \cite{Sato:2011qr} at the $1 \%$ level.

\begin{figure}
\centerline{\includegraphics[width = 17cm,keepaspectratio=true]{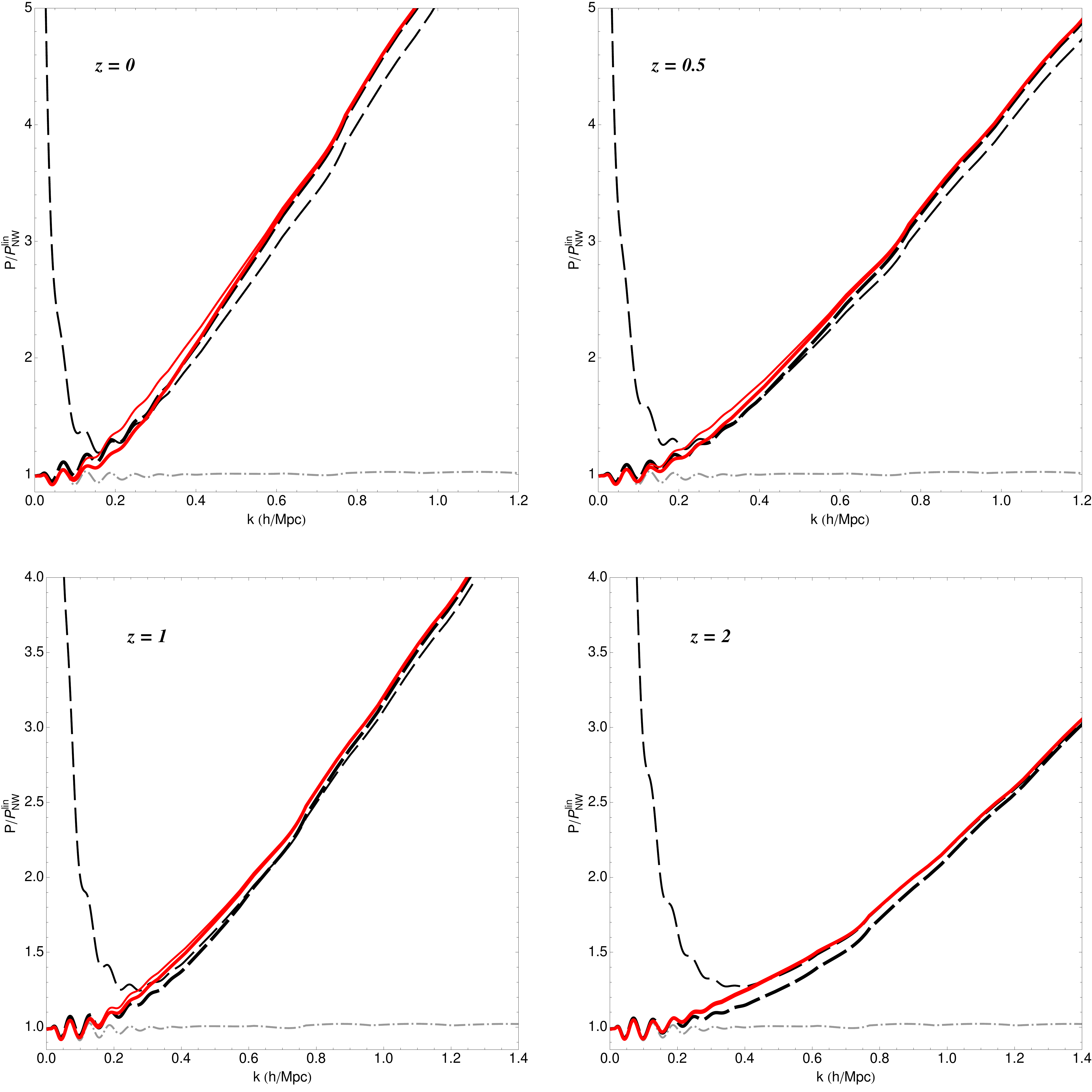}}
\caption{The PS at large $k$ from the evolution equation in various approximations for the mode-mode coupling part in the third line of eq.~\re{TReik} and at different redshifts. The thick black dashed line corresponds to using eq.~\re{phig}, the thin red solid line to the improved $\tilde{\Phi}_{ab}$ of eq.~\re{largephitnew} with no filter imposed in the second term, and the thick red solid line to using eq.~\re{largephitnew} with the power-law filter function of eq.~\re{filter} with $\bar{k}=0.2\,\mathrm{h/Mpc}$ (see \ref{AB} for explicit formulae).
Also shown are the large $y$ attractor solution of eq.~\re{approxasy} (thin black dashed line), and linear PT (grey dash-dotted line).  }
\label{PS_Large}
\end{figure}

\begin{figure}
\centerline{\includegraphics[width = 17cm,keepaspectratio=true]{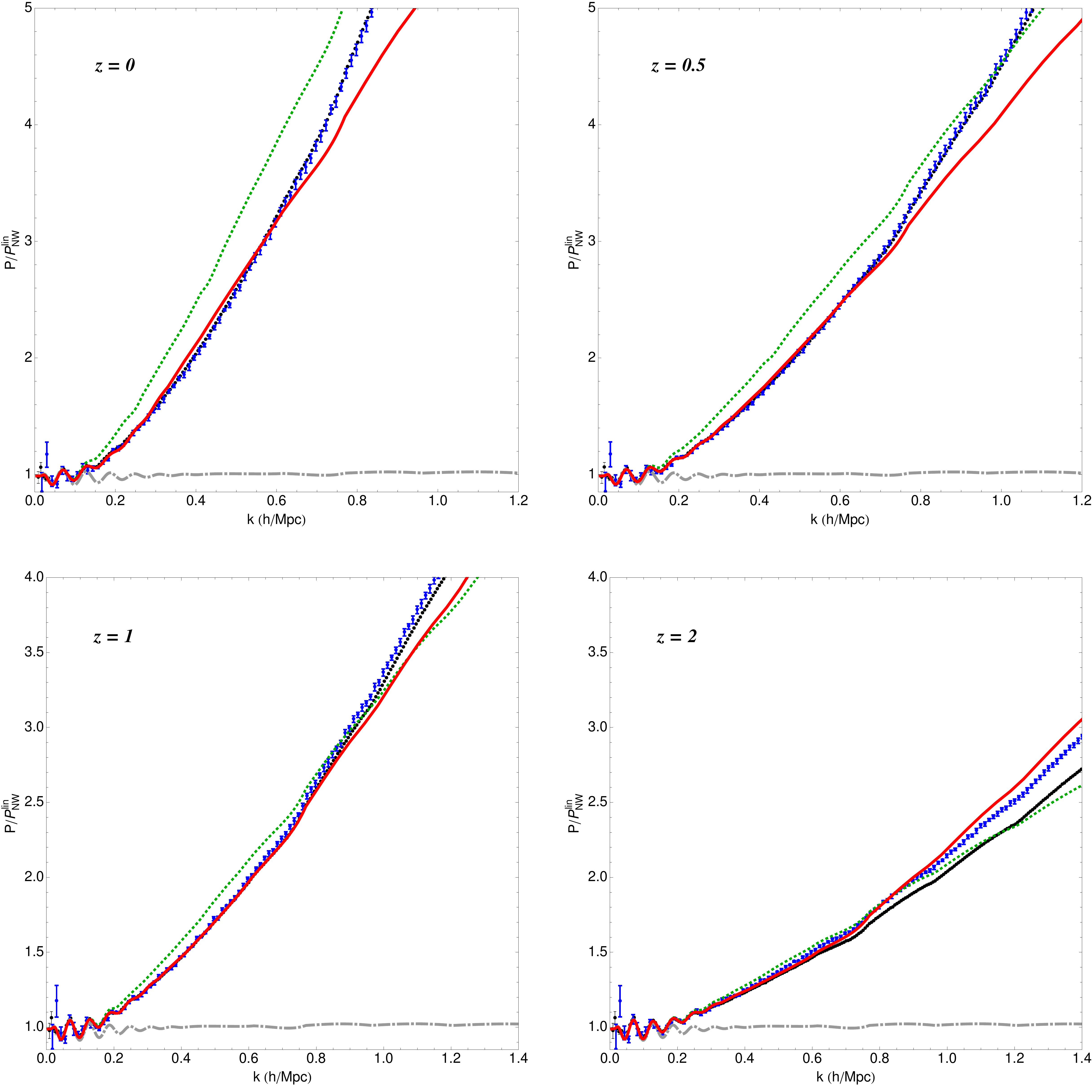}}
\caption{Comparison between the evolution equation discussed in this paper (red solid line, see \ref{AB} for explicit formulae), and the N-body simulations of Sato and Matsubara \cite{Sato:2011qr}: black dots are for the large volume simulation ($L=1000\,\mathrm{Mpc/h}$), blue ones for the small volume one ($L=500\,\mathrm{Mpc/h}$). Also shown are linear PT (grey dash-dotted line) and 1-loop PT (green dotted line).}
\label{PS_Large_Comp}
\end{figure}

In fig.~\ref{PS_Large}, we plot, in an extended $k$ range, the results obtained as in the red line of fig.~\ref{PS_BAO} with (thick red) and without (thin red) the filter function of eq.~\re{filter}. We see that, as expected, the effect of the filter vanishes at large $k$ (where the filter goes to unity) and at increasing redshift, where the relative weight of the second term in eq.~\re{largephitnew} compared to the first one becomes less relevant. We also plot (thick black dashed line) the result obtained by using the large $k$  expression of eq.~\re{phie} in eq.~\re{largephitnew}. We also show the analytic expression for the attractor solution found in eq.~\re{approxasy} (thin black dashed line).

The comparison with the nonlinear PS from the N-body simulations at large $k$ is given in fig.~\ref{PS_Large_Comp}. In order to gauge the performance of the N-body results at large $k$, Sato and Matsubara performed runs with two different volumes, a large one ($L_{box}=1000\,\mathrm{Mpc/h}$) and a small one ($L_{box}=500\,\mathrm{Mpc/h}$), which we plot with black and blue points, respectively. We notice that the two sets of N-body data are practically overlapped for $z<1$, where we also checked that they agree with the Coyote emulator, but diverge significantly at $z\ge2$ for $k\agt  0.8\, \mathrm{h/Mpc}$. These transient effects therefore prevent us from considering the comparison for higher redshifts and scales, however, the trend from $z=0$ up to $z=2$ clearly shows a progressive improvement, as it should. Quantitatively, at $z=1$ we measure an agreement at the $1\%$ level between our results and the N-body simulations up to $k=  0.8\, \mathrm{h/Mpc}$ and at $2\%$ on the same range of scales for $z=0.5$.

\section{Discussion and conclusions}
\label{discussion}
The scheme discussed in this paper presents three main advantages with respect to alternative approaches to the nonlinear PS in the $k\alt 1 \;\mathrm{h\,Mpc}^{-1}$ scale, including N-body simulations: the accuracy (already discussed in the previous section), the computational speed, and the range of cosmologies that can be dealt with.
The computational time is the same as that required by a standard PT computation at 1-loop. Indeed, concerning momentum integrations, to obtain the three independent components of the PS, one needs to perform two one-dimensional integrals, eqs.~\re{HH1} and \re{HH2}, and the three two-dimensional ones of eq.~\re{phicomp}. Exactly the same integrals enter the 1-loop expression for the PS, eq.~\re{PS1loop}. In addition, the solution of our evolution equation requires only one more integration in $\eta$, which however takes a time of the order of a second. Therefore,  any point in $k$ takes at most a few seconds to be evaluated.

The extension from $\Lambda$CDM cosmology to more general ones is also greatly eased in this approach. In principle, all the cosmologies in which the fluid equations of eq.~\re{Euler} are modified only in the linear terms, can be taken into account by a modification of the $\Omega_{ab}$ matrices in \re{bigomega}, as it was discussed in \cite{Pietroni08}. Cosmologies of this type include, for instance, those with massive neutrinos \cite{STT08,LMPR09,SaitoII}, modifications of gravity of the scalar-tensor/f(R) type \cite{Saracco:2009df}, or Dark Energy models with a non-relativistic sound speed \cite{Anselmi:2011ef}. Non-gaussian initial conditions can also be taken into account by the inclusion of new vertices in the diagrammatic rules of fig.~\ref{FEYNMAN} and their impact on the propagator and on the PS can be analyzed \cite{Bartolo:2009rb,Bernardeau:2010md}.

Evolution equations for the nonlinear PS have been proposed also before \cite{McD06,Taruya2007, Pietroni08,TaruyaIII}. The main step  forward provided by the present analysis is, besides the computational speed, the improved treatment of the large $k$ limit, in particular for the mode-mode coupling term. Indeed, while the gaussian damping of the contribution proportional to the linear PS, namely the first term in eq.~\re{fullP}, is reproduced satisfactorily well by most resummed PT methods, the remaining part, that containing the mode-mixing function  $\Phi_{ab}$ is usually included only at lowest order in PT, adding at most a few corrections. As a result, the improvement of these methods over the 1-loop results was not able to go beyond the BAO scale, as it was discussed in detail for the Time Renormalization Group approach of ref.~\cite{Pietroni08} in    \cite{Audren:2011ne, Juergens:2012fk}.

A different strategy to resum PT corrections at all orders is provided by the approach of \cite{Bernardeau:2008fa}. In this context, it was shown that the exact CS resummation for the propagator holds also for  the larger class of multipoint propagators. The nonlinear PS can be expressed as an infinite sum over squares of multi-point propagators, however  the kinematical regime for which the exact resummation was proved is different from the one considered here, namely, the one in which only one internal PS is at large $k$, the other ones being soft. Our results, and their good comparison with N-body simulations, indicate that the relevant kinematical configurations in the intermediate $k$ regime are still given by those loops in which most of the external momentum is carried by a single internal leg, and that the dominant diagrams are still of the chain type, with some amount of vertex renormalization neglected, as detailed in subsection \ref{intk}.

 Our results depend on the fact that $\Lambda$CDM-like PS's are well behaved as $k\to0$ and have an intrinsic length scale which is of the order of $\sigma_v e^\eta$. This makes a separation between `hard' and `soft' modes possible, and, due to the $k/q$ enhancement of the vertex functions, singles out the class of chain diagrams as the dominant one at large $k$. A very different approach would be needed to take into account, for instance, pure power-law PS's. In this case the dominant mode coupling at a scale $k$ would involve modes of $O(k)$ as well. Fortunately such cases are nowadays only of academic interest.

The approach presented in this paper suffers from an intrinsic physical limitation, namely, the neglect of velocity dispersion and all higher order moments of the particle distribution function, which is at the basis of the derivation of eqs.~\re{Euler} from the Vlasov equation. This ``single stream approximation" is known to hold at large scales and high redshifts, but it was estimated to fail at the percent level in the BAO range at $z\to 0$ \cite{Pueblas:2008uv,Valageas:2010rx}. The comparison of Sect.~\ref{results} between our results and N-body simulations exhibits the same trend. It will be interesting to investigate further the origin of this excellent agreement. 

The access to the multi-stream regime is precluded by construction to all forms of (improved) PT, including this one. A way to incorporate such effects in semi-analytical methods was recently proposed in \cite{Pietroni:2011iz}, in which the feeding of the multi stream at small scales on the more perturbative intermediate scales was described in terms of effective source terms. The inclusion of such effects in the present approach will be studied elsewhere.

For the time being, the increase of the maximum $k$ at which the nonlinear PS can be computed reliably, provided by our approach, opens the way to interesting cosmological applications, allowing a tighter extraction of cosmological parameters from the LSS relevant for topics such as the measurement of the acoustic scale from BAO's, the limit on the neutrino mass scale, and possibly, cosmic shear. 

\section*{Acknowledgments}
We thank M. Sato and T. Matsubara as well as M. White and J. Carslon for providing the simulations data.
We thank M. Crocce for useful discussions and SA would like to acknowledge M. Fael for teaching him the C-coding and the MC integration techniques. SA thanks the Institut de Ci\`encies de l'Espai (IEEC/ICE), Barcelona, Spain, for hospitality, and ``Fondazione Angelo Della Riccia'' and ``Fondazione Ing.~Aldo Gini'' for financial support. MP acknowledges partial support from the  European Union FP7  
ITN INVISIBLES (Marie Curie Actions, PITN- GA-2011- 289442).
\appendix
\section{}
\label{eRPT}
In this appendix we will discuss a scheme to do renormalized cosmological perturbation theory in a way that closely resembles the RPT of \cite{RPTa,RPTb} but gives a systematic way to compute higher loop orders. In particular, we will stress the crucial role of counter terms, which should be added to the diagrammatic rules in order to avoid overcounting of the contributions. Since the `tree level' in this framework coincides, in the large momentum limit, with the eikonal approximation of  \cite{Bernardeau:2011vy, Bernardeau:2012aq} we will refer to this scheme es eRPT.

The root of RPT as defined in \cite{RPTa,RPTb} is to use, as tree propagator, the resummed one {\it i.e.} the one decaying as $\exp \big[-k^2 \sigma_v^2 (e^\eta-e^{\etap})^2/2\big]$ at large $k$, instead of the linear one, $g_{ab}(\eta,\etap)$. Since the low-$k$ limit of the resummed propagator does not match with 1-loop standard PT, an interpolation procedure is usually advocated in order to have a well behaved tree propagator, going to the 1-loop  (in standard PD) one as $k\to 0$.

An alternative approach is to define as the tree level propagator, for all values of $k$, the resummed propagator
\beq
G^{eik}_{ab}(k;\eta,\etap) = g_{ab}(\eta,\etap)\;\exp \big[-k^2 \sigma_v^2 \frac{ (e^\eta-e^{\etap})^2}{2}\big]\,,
\label{ge}
\eeq
and as tree level PS, 
\beq
P^{eik}_{ab}(k;\eta,\etap)= P^0(k) u_a u_b  \;\exp \big[-k^2 \sigma_v^2 \frac{ (e^\eta-e^{\etap})^2}{2}\big]\,,
\label{pe}
\eeq
where $P^0(k)$ is the linear PS in standard PT and we have assumed growing mode (and adiabatic) initial conditions.
The above quantities are obtained starting from the linear ones in standard PT and correcting them by adding all possible  `chain-diagrams', {\it i.e.} diagrams in which `soft' PS at momenta $q_i$ are attached to the hard line at momentum $k$, such that $q_i\ll k$. The results above follow as a consequence of the property of the vertex function, eq.~(\ref{vertice}),
\beq
\gamma_{abc}(\bk,-\bk+\bq,-\bq)\,u_c \to \delta_{ab}\frac{1}{2} \frac{\bk\cdot\bq}{q^2} \,,\qquad\qquad\qquad \mathrm{for}\; k \gg q\,,
\label{lkv}
\eeq
and of the composition property of linear propagator,
\beq
g_{ab}(s,s')g_{bc}(s',s'')=\theta(s-s')\theta(s'-s'') g_{ac}(s,s'')\,.
\label{cp}
\eeq
Notice, in particular, as also discussed in the text, that the equal-time PS is equal to the linear one.

These expressions are exact in the deep eikonal limit, that is for very large momentum $k$ (neglecting multi streaming) {\em and} under the assumption that the cosmological scenario presents a clear hierarchy of scales allowing us to set apart `soft' from `hard' modes. Corrections to this idealized situation come from loop corrections, once they are properly taken into account. This can be done by using the path-integral formulation introduced in \cite{MP07b} (see also \cite{Anselmi:2010fs}). In that paper, it was shown that standard PT is equivalent to an expansion in powers of the vertex function $\gamma_{abc}$ of the generating functional
 \beqra
\label{genf}
&&Z[J_a,\, K_b] 
=\nonumber\\
&& \int {\cal D}  \vp_a {\cal D} \chi_b
\exp \biggl\{-\half \int d\eta d\eta'   \chi_a(\eta) \chi_b(\etap) P^0(k) u_a u_b 
\delta(\eta-\etain) \delta(\eta'-\etain) \nonumber\\
&+& 
i \int d\eta \left[ \chi_a g^{-1}_{ab} \vp_b  -  
e^\eta\,\gamma_{abc} \chi_a \vp_b \vp_c + 
J_a \vp_a +  K_b \chi_b\right]\biggr\} \,, 
\label{GENFUN}
\eeqra
 where $J_{a}$ and $K_{b}$ are sources for $\varphi_{a}$ and $\chi_{b}$ respectively, and we have omitted the momentum dependence. In deriving the above expression we have assumed Gaussian initial conditions. Non-Gaussian initial conditions can be taken into account by including a non-vanishing bispectrum, trispectrum, etc., in the first line of Eq.~(\ref{GENFUN}).
Derivatives of Eq.~(\ref{GENFUN}) w.r.t. the sources $J_a$ and $K_b$ give all the possible statistical correlators involving the fields $\vp_a$ and $\chi_b$, such as the nonlinear PS and nonlinear propagator, given by 
\beqra
&&\left. \frac{1}{Z}\frac{\delta^2 Z}{\delta J_a(\bk,\eta)\, \delta J_b(\bk',\etap)}\right|_{J_a,\,K_b=0} 
= - \, \delta_D({\bk}+{\bk}') P_{ab}(k;\eta,\etap)\,,\nonumber\\
&&\left.\frac{1}{Z}\frac{\delta^2 Z}{\delta J_a(\bk,\eta)\, \delta K_b(\bk',\etap)}\right|_{J_a,\,K_b=0} 
= -i\, \delta_D({\bk}+{\bk}') G_{ab}(k;\eta,\etap)\,,
\eeqra
respectively.

In order to pass from standard PT to eRPT, without modifying the full dynamical content of the generating functional,  we simply add and subtract the quadratic expression
\beq
\int d\eta d\eta'   \bigg[-\half  \chi_a(\eta)  \Phi^{eik}_{ab}(\eta,\etap) \chi_b(\etap)  -i  \chi_a(\eta) \Sigma^{eik}_{ab}(\eta,\etap) \vp_b(\etap)  \bigg]\,,
\eeq
to the exponent of eq.~(\ref{GENFUN}). Then, we include the added term in the new ``free'' ({\it i.e.} quadratic) part of the action, whereas the subtracted term goes in the new ``interaction'' part, which also includes the trilinear vertex at the third line of eq.~\re{GENFUN}. Of course, since we have just added and subtracted the same term, standard PT and eRPT have the same dynamical content, and their results  fully coincide at infinite loop order. 

The `self-energy' and `mode coupling' function in the eikonal limit, $\Sigma^{eik}_{ad}(k;\eta,\etap)$ $\Phi^{eik}_{ad}(k;\eta,\etap)$, can be represented by diagrams in which the hard line, carrying momentum of order $k$, is corrected by attaching  soft PS in all possible ways such that the final diagram is one-particle-irreducible (1PI), that is, it cannot be cut into two disjoint pieces by cutting a single line. In terms of eikonal propagators, $PS$, and renormalized vertices, they are given by the 1-loop diagram of fig.~\ref{geik}, and by the 1- and 2-loop diagrams at the RHS of fig.~\ref{PhiGeik}, respectively, in which the tree rightmost vertex is replaced by a renormalized one (thick dot). In all these diagrams the propagator at the right, connecting $s'$ to $\eta$, has to be cut-off.

Setting to zero the new interaction term, the path integral can be performed analytically, to give
\beqra
&&Z_0^{eRPT}[J_a,\, K_b]=\exp\left\{
-  \int d\eta d\etap \left[\half J_a(\eta) 
P^{eik}_{ab}(k;\eta,\etap) J_b(\etap)\right.
\right.\nonumber \\
&&\qquad\qquad\qquad\qquad \qquad\left. \left.+i \,J_a(\eta) 
G^{eik}_{ab}(k;\eta,\etap) 
K_b(\etap)\right]\right\}\,,
\label{zfree}
\eeqra
where $G^{eik}_{ab}$ and $P^{eik}_{ab}$ are given in eqs.~\re{ge} and \re{pe}, respectively, and are related to $ \Sigma^{eik}_{ab}$ and $ \Phi^{eik}_{ab}$ by 
\beqra 
G^{eik}_{ab}(k;\eta,\etap) &=& \big[ g^{-1}-  \Sigma^{eik}  \big]_{ab}^{-1}(k;\eta,\etap)\nonumber\\
&=& g_{ab}(\eta,\etap) + \int d s ds'\,  g_{ac}(\eta,s)\Sigma^{eik}_{cd}(k;s,s')  G^{eik}_{db}(k;s',\etap)\,,
\eeqra
and
\beqra
&& P^{eik}_{ab}(k;\eta,\etap) = G^{eik}_{ac}(k;\eta,\etain) G^{eik}_{bd}(k;\etap,\etain) P^0(k)u_a u_b \nonumber\\
&&\quad\qquad\qquad+\int d s ds'\,  G^{eik}_{ac}(k;\eta,s) G^{eik}_{bd}(k;\etap,s') \Phi^{eik}_{cd}(k;s,s')\,.
\label{pphi}
\eeqra
\begin{figure}
\centerline{\includegraphics[width = 15cm,keepaspectratio=true]{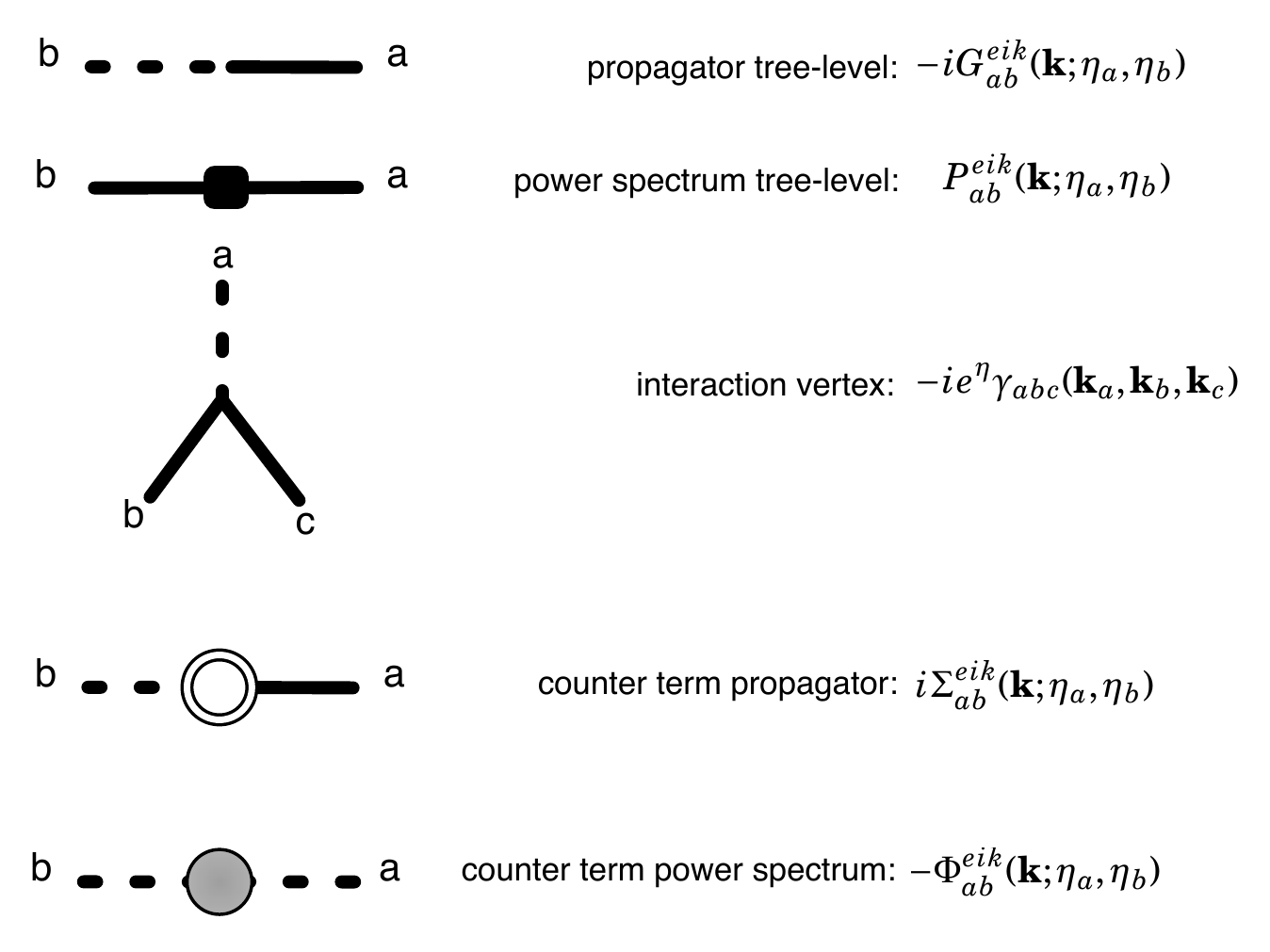}}
\caption{The Feynman rules for eRPT.}
\label{feRPT}
\end{figure}
The new expansion can be expressed diagrammatically as illustrated in fig.~\ref{eRPT}. Besides the eikonal propagator and PS given above, in the interaction sector we have the usual trilinear interaction and two new counterterms, $i \Sigma^{eik}_{ab}(k;s,s') $ and $-  \Phi^{eik}_{ab}(k;s,s')$. These counterterms avoid overcounting of the contributions already included in the resummed propagator and PS.

\begin{figure}
\centerline{\includegraphics[width = 15cm,keepaspectratio=true]{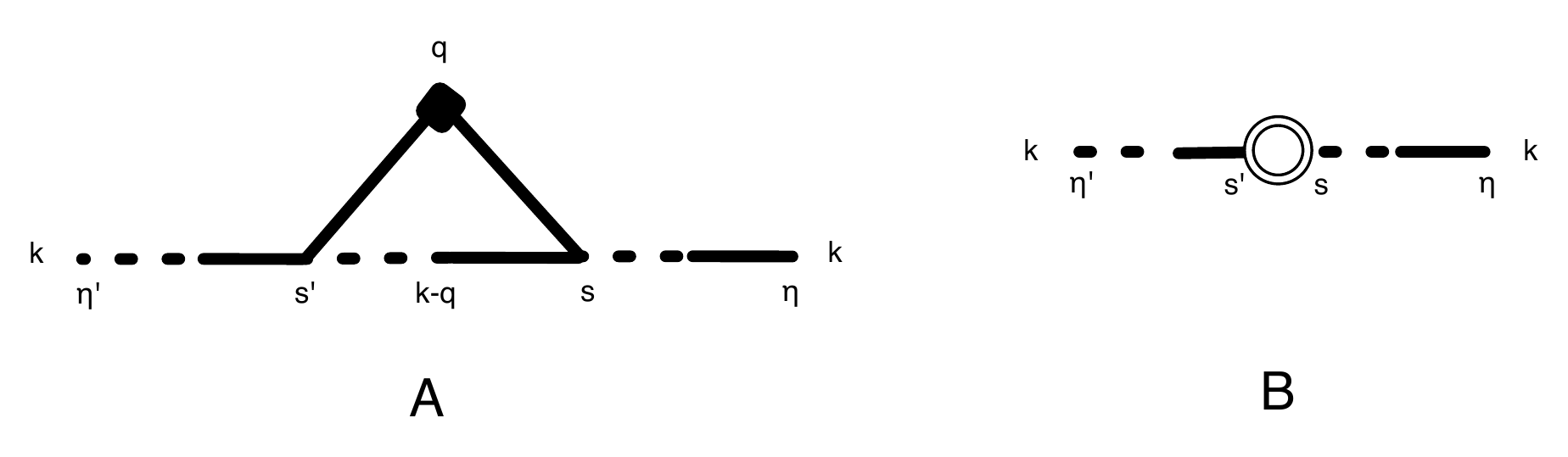}}
\caption{The first non-trivial corrections to $G_{ab}^{eik}$ in eRPT.}
\label{1lGeRPT}
\end{figure}

To see how eRPT works, we consider  the propagator up to 1-loop order. The corrections to the tree level expression, \re{ge}, are given by the 1-loop contribution of fig.~\ref{1lGeRPT}
\beq
 \int ds\,ds' G^{eik}_{ac}(k;\eta,s)\Sigma^{eRPT, (1)}_{ch}(k;s,s')G^{eik}_{hb}(k;s',\etap)\,, 
\label{1lerpt}
\eeq
where
\beqra
&&\Sigma^{eRPT, (1)}_{ch}(k;s,s') = 4 \,e^{s+s'}  \int d^3\bq\;\gamma_{cde}(k,q,|\bk-\bq|) G^{eik}_{ef}(|\bk-\bq|;s,s') \times\nonumber\\
&&\quad\qquad\qquad\qquad\qquad P^{eik}_{dg}(k;s,s')\gamma_{fgh}(|\bk-\bq|,q,k)
\eeqra
and by the contribution of the counterterm,
\beqra
&&-\int ds\,ds' \;G^{eik}_{ac}(k;\eta,s) \Sigma^{eik}_{cd}(k;s,s')G^{eik}_{db}(k;s',\etap)  \nonumber\\ 
&&= k^2 \sigma_v^2 \int ds\, e^s(e^s-e^{\etap}) G_{ac}^{eik}(k;\eta,s) G_{cb}^{eik}(k;s,\etap) \nonumber\\
&&= \sqrt{\pi}\, \frac{k \sigma_v}{2} (e^\eta-e^{\etap})\, \mathrm{Erf}\Big[ \frac{k \sigma_v}{2} (e^\eta-e^{\etap}) \Big] \exp\Big[ -\frac{k^2 \sigma_v^2}{4}( e^\eta-e^{\etap})^2\Big]\,g_{ab}(\eta,\etap)\,.\nonumber\\
&&
\label{ct}
\eeqra
Notice that the 1-loop term, eq.~\re{1lerpt}, contains the resummed propagator and PS, but tree level vertices. The latter appear in their exact  form, that is, not in the eikonal limit of eq.~\re{lkv}.
In the small $k$ limit the tree level propagator, eq.~\re{ge} goes as
\beq
g_{ab}(\eta,\etap)\Big[1-k^2 \sigma_v^2 \frac{(e^\eta-e^{\etap})^2}{2} + O(k^4)\Big]\,.
\eeq
The $O(k^2)$ term is exactly canceled by the counter term in \re{ct}, which gives
\beq
k^2 \sigma_v^2 \frac{(e^\eta-e^{\etap})^2}{2} g_{ab}(\eta,\etap)+ O(k^4)\,,
\eeq
so that, summing up eqs.~\re{ge}, \re{1lerpt}, and \re{ct}, one gets 
\beq
g_{ab}(\eta,\etap)+ \int ds\,ds' g_{ac}(\eta,s)\Sigma_{ch}^{(1)}(k;s,s') g_{hb}(s',\etap) +O(k^4)\,,
\eeq
with
\beqra
&&\Sigma_{ch}^{(1)}(k;s,s') =4\,e^{s+s'}\int d^3\bq\;  \gamma_{cde}(k,q,|\bk-\bq|) g_{ef}(s,s') \times\nonumber\\
&&\quad\qquad\qquad\qquad\qquad P^0(k) \,u_d u_g\,\gamma_{fgh}(|\bk-\bq|,q,k)\,,
\eeqra
that is, we obtain the propagator at 1-loop {\em in standard PT}, $g_{ab}(\eta,\etap)+\Delta g^{(1)}_{ab}(\eta,\etap)$. In order to get the $O(k^4)$ terms of standard PT one has to include the 2-loop diagrams of eRPT -- not only the chain-like ones -- and up to two insertions of the $ \Sigma^{eik}_{ab}$ counter terms. Indeed, the single counter term insertion in \re{ct}, expanded up to $O(k^4)$, gives $-1/6 \; k^4 \sigma_v^4(e^\eta-e^{\etap})^4$, while the double insertion gives
\beqra
&&\int ds\, ds' \,ds''\, ds''' \;G^{eik}_{ac}(k;\eta,s) \Sigma^{eik}_{cd}(k;s,s')G^{eik}_{de}(k;s',s'')  \nonumber\\
&&\qquad\qquad\qquad \qquad \qquad \times \Sigma^{eik}_{ef}(k;s'',s''')G^{eik}_{fb}(k;s'''',\etap) \nonumber\\
&& = \frac{k^4 \sigma_v^4}{4!} (e^\eta-e^{\etap})^4 g_{ab}(\eta,\etap) + O(k^6)\,,
\eeqra
so that the contribution from counterterms  exactly cancels the $ k^4 \sigma_v^4 (e^\eta-e^{\etap})^4/8 $ term coming from the expansion of $G_{ab}^{elk}$.

 Notice that the appearance of the tree level vertex in its complete momentum dependence ({\it i.e.} not in the approximated form \re{lkv}) is essential in order to recover the correct momentum dependence of the propagator, or, analogously, of the PS. 

The correction to the tree level propagator \re{ge}, {\it i.e.} the sum of \re{1lerpt} and \re{ct} amounts to taking one of the soft PS insertion in the chain diagrams in the eikonal limit, and to replacing it with an insertion in which the vertices are at the tree level in their full form (instead of being fully renormalized, but in the eikonal approximation), and the inserted PS is also renormalized, {\it i.e.} it carries gaussian damping factors, as in \re{pe}. In other words, the contribution in \re{1lerpt} represents a 1-loop ``super-chain'' diagram, in which, compared to the usual 1-loop chain diagram, all the propagators and  the internal PS are renormalized. 

Differently from standard PT, at large $k$ the different orders in this expansion do not diverge as some power law in $k$, but are always damped by the gaussian decay of the external eikonal propagators. In the large $k$ limit the vertices can be approximated as in \re{lkv} and the summation of the super-chain diagrams at all orders gives rise to a super-resummed propagator. In this resummation the counter terms should be properly taken into account in order to avoid over counting of the standard chain diagrams already included at the tree level. At intermediate scales, the full momentum dependence of the vertices, and also non-chain diagrams should be taken into account to properly compute the propagator away from the eikonal limit, as we have done up to $O(k^4)$ corrections.

\begin{figure}
\centerline{\includegraphics[width = 15cm,keepaspectratio=true]{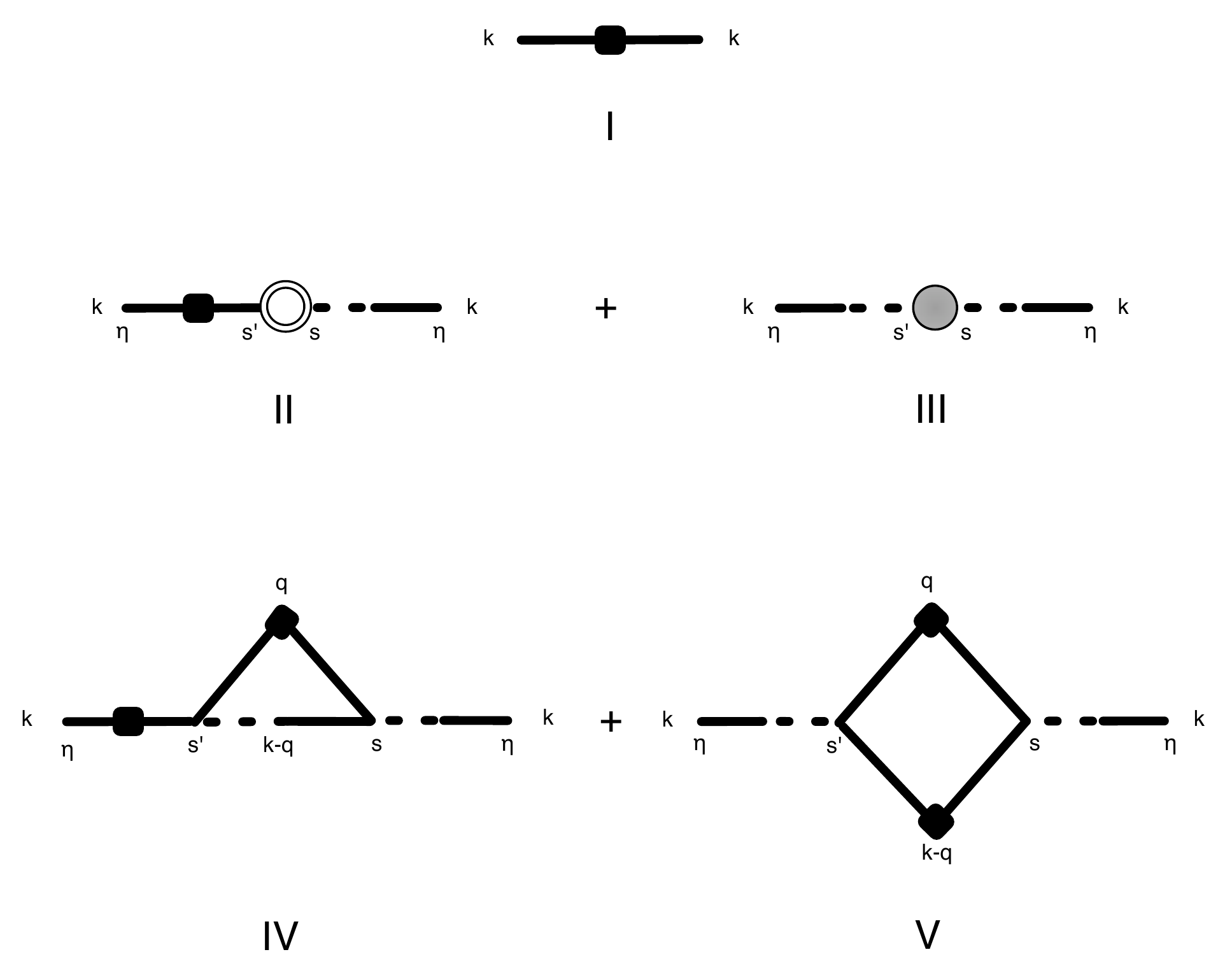}}
\caption{The first non-trivial corrections to $P_{ab}^{eik}$ in eRPT.}
\label{1lPSeRPT}
\end{figure}

We now consider the PS. At 1-loop in eRPT we need to consider the diagrams in fig.~\ref{1lPSeRPT}, where also the counterterm  $\Phi^{elk}_{ab}$ now appears. Also in this case, the 1-loop PS in standard PT is reproduced, at low $k$ up to $O(k^4)$ terms. The role of the external propagator corrections, and of the corresponding counterterms
 $\Sigma^{eik}_{ab}$ is analogous to what we have just discussed for the propagator. As for the mode-coupling part, the sum of the tree level PS of eqs.~\re{pe} (see also \re{pphi})
and of the counterterm gives the first line in eq.~\re{pphi}, which is doubly gaussian suppressed, and plays therefore a subleading role at intermediate scales. We are therefore left with the 1-loop (in eRPT) contribution
\beq
\int d s ds'\,  G^{eik}_{ac}(k;\eta,s) G^{eik}_{bd}(k;\etap,s') \Phi^{eRPT, (1)}_{cd}(k;s,s')\,,
\eeq
where
\beqra
&&\Phi^{eRPT, (1)}_{cd}(k;s,s') = \nonumber\\
&& 2\, e^{s+s'} \int d^3 \bq \gamma_{acd}(k,q,|\bk-\bq|) \gamma_{bef}(k,q,|\bk-\bq|) P^{eik}_{ce}(q;s,s')P^{eik}_{df}(|\bk-\bq|;s,s')\,,\nonumber\\
\label{phi1erpt}
\eeqra
Which we will use in eq.~\re{phie1} in its approximated form valid for $k\gg q$.

\section{}
\label{3L}
In this appendix we prove  eq.~\re{Teik}. First, we notice that, as proved  in \cite{Anselmi:2010fs}, we have
\beqra
\Delta G^{eik,\,(l)}_{ac}(k;\eta,s)  &=& G^{eik,\,(l-1)}_{{\bf a}c}(k;\eta,s)\,  \int_s^\eta ds' \,\Sigma^{eik, \,(1)}_{{\bf a}e}(k; \eta, s')u_e\,,\nonumber\\
&=& H_{\bf{a}}(k;\eta,s) \,G^{eik,\,(l-1)}_{{\bf a}c}(k;\eta,s)\,,
\label{dg}
\eeqra
where the loop indices refer to standard PT, not to the eRPT described in \ref{eRPT}.
Then, in order to reproduce  \re{Tstartv}, we have to show that a similar expression, but with the first factor replaced by $\int_{\etain}^s ds' \,\Sigma^{eik, \,(1)}_{{\bf a}e}(k; \eta, s')u_e$, can be extracted from the third line in \re{Teik}. Indeed, we can write 
\beq
\Big[\Phi^{eik}_{ad}(k;\eta,s') G^{eik}_{bd}(k;\,\eta,s') \Big]^{(n)} = \sum_{j=1}^n \Phi^{eik,\,(j)}_{ad}(k;\eta,s') \,G^{eik,\,(n-j)}_{bd}(k;\,\eta,s') \,.
\eeq
The $j$-th order contribution to the mode coupling function in the eikonal limit, $\Phi^{eik,\,(j)}_{ad}(k;\eta,s')$ can be represented by diagrams in which the line containing the hard linear PS, carrying momentum of order $k$, is corrected by attaching $j$ soft PS in all possible ways such that the final diagram is one-particle-irreducible, that is, it cannot be cut into two disjoint pieces by cutting a single line.
Then, we do the splitting
\beq
\Phi^{eik,\,(j)}_{ad}(k;\eta,s') = \Phi^{eik,\,L,\,(j)}_{ad}(k;\eta,s') + \Delta  \Phi^{eik,\,L,\,(j)}_{ad}(k;\eta,s') \,, 
\eeq
where $ \Phi^{eik,\,L,\,(j)}_{ad}(k;\eta,s')$ contains all the diagrams in which the soft PS attached to the left end, here identified by index ``a" and time ``$\eta$", has the other leg attached to the right of the hard PS, and $ \Delta  \Phi^{eik,\,L,\,(j)}_{ad}(k;\eta,s')$ contains all the other diagrams. Equivalently, the diagrams belonging to $ \Phi^{eik,\,L,\,(j)}_{ad}(k;\eta,s')$ have the leftmost vertex at the tree level.

\begin{figure}
\centerline{\includegraphics[width = 15cm,keepaspectratio=true]{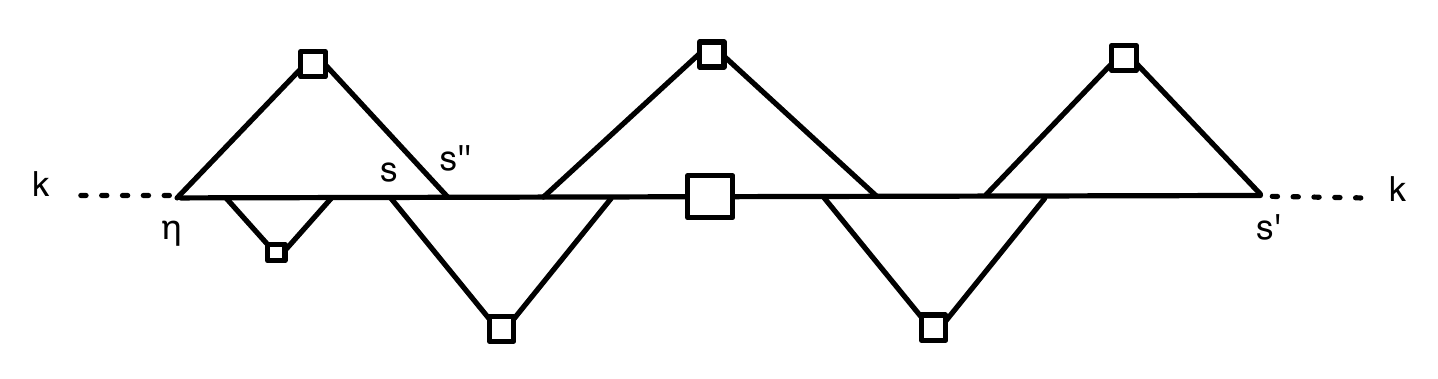}}
\caption{A generic contribution to  $\Delta  \Phi^{eik,\,L,\,(j)}_{ad}(k;\eta,s')$.}
\label{DPL}
\end{figure}

A generic diagram contributing to  $ \Delta  \Phi^{eik,\,L,\,(j)}_{ad}(k;\eta,s')$ is given in fig.~\ref{DPL}. In the eikonal limit the soft PS attached to the leftmost end gives a factorized contribution of the form $\Sigma^{eik, \,(1)}_{ae}(k; \eta, s'')\,u_e=-u_a \,k^2 \sigma_v^2 e^{\eta+s''}$. This is a consequence of the property of the elementary vertex of eq.~(\ref{lkv}), and of the composition property of linear propagator, eq~\re{cp}.

We are left with a contribution to the $(j-1)$-th order quantity
\beq
\int_{\etain}^{\eta} ds \Big[G^{eik}_{ac}(k;\eta,s)\Phi^{eik}_{cd}(k, s,s')\Big]^{(j-1)}\,,
\label{gphi}
\eeq
where the time $s$ corresponds to the point where the 1PI function $\Phi^{eik}_{cd}(k, s,s')$ is joined to the connected one, $G^{eik}_{ac}(k;\eta,s)$. Notice, in particular, that the time argument $s''$ of the factorized $\Sigma^{eik, \,(1)}$-function is, in any case $\le s$, since, otherwise, the corresponding diagram for $ \Delta  \Phi^{eik,\,L,\,(j)}_{ad}(k;\eta,s')$  would not be 1PI.
Now, summing up all the contributions to  $\Delta  \Phi^{eik,\,L,\,(j)}_{ad}(k;\eta,s')$ containing the {\em same} particular diagram contributing to eq.~\re{gphi} with the end of $\Sigma^{eik, \,(1)}$ at $s''$ attached in all possible ways, gives 
\beq
 \int_{\etain}^{\eta} ds\; \int_{\etain}^s ds' \,\Sigma^{eik, \,(1)}_{{\bf a}e}(k; \eta, s')u_e \;  \Big[G^{eik}_{{\bf a}c}(k;\eta,s)\Phi^{eik}_{cd}(k, s,s')\Big]^{(j-1)}\,.
 \label{f2}
\eeq
Inserting \re{dg} in the first line of \re{Teik} and \re{f2} (and the corresponding contribution for $a\leftrightarrow b$) in the third line, proves the equality.

\section{}
\label{AA}
In this appendix we will give the explicit expressions of the 1PI functions $\Sigma_{ab}^{(1)}(k;\eta,\etap)$ and $\Phi_{ab}^{(1)}(k;\eta,\etap)$ computed at 1-loop. The general expression for $\Sigma_{ab}^{(1)}(k;\eta,\etap)$ is given by
\beqra
&&\Sigma_{ab}^{(1)}(k;\eta,\etap) =\nonumber\\
&&\quad  4 e^{\eta+\etap} \int d^3 q \gamma_{acd}(\bk,-\bq,\bq-\bk) u_c P^0(q) u_e \gamma_{feb}(\bk-\bq,\bq,-\bk) g_{df}(\eta,\etap)\,,
\label{sigform}
\eeqra
which, using the expressions for the vertices and the linear propagator in eqs. \re{vertice}, \re{proplin}, and performing the angular integration, gives
\beqra
&&\Sigma_{11}^{(1)}(k;\eta,\etap)=-e^{-3/2\,\eta+\etap}\frac{k^3\,\pi}{15} \int dr\,P^0(kr) \Bigg[3 e^{5/2\,\eta}(1-3r^2)\nonumber\\
&&\qquad\qquad+e^{5/2\,\etap}(17+9r^2)+\frac{9}{2r}\left(e^{5/2\,\eta}-e^{5/2\,\etap}\right)(r^2-1)^2\log\left|\frac{1+r}{1-r}\right|\Bigg]\,,\nonumber\\
&&\Sigma_{12}^{(1)}(k;\eta,\etap)=-e^{-3/2\,\eta+\etap}\frac{k^3\,\pi}{15}  (e^{5/2\eta}-e^{5/2\etap}) \int dr\,P^0(kr) \Bigg[8-15 r^2+ 9 r^4 \nonumber\\
&&\qquad  -\frac{9 r}{2} (r^2-1)^2\log\left|\frac{1+r}{1-r}\right|\Bigg]\,,\nonumber\\
&&\Sigma_{21}^{(1)}(k;\eta,\etap)=-e^{-3/2\,\eta+\etap}\frac{3\, k^3\,\pi}{5} (e^{5/2\,\eta}-e^{5/2\,\etap})\int dr\,P^0(kr) \Bigg[\frac{1}{r^2}(3 r^2-1)\nonumber\\
&&\qquad\quad\qquad+\frac{1}{2r^3}(r^2-1)^2\log\left|\frac{1+r}{1-r}\right|\Bigg]\,,\nonumber\\
&&\Sigma_{22}^{(1)}(k;\eta,\etap)=-e^{-3/2\,\eta+\etap}\frac{k^3\,\pi}{15} \int dr\,P^0(kr) \Bigg[3e^{5/2\,\etap} (1-3r^2)+e^{5/2\,\eta} (17+9r^2)\nonumber\\
&&\qquad\quad\qquad-(e^{5/2\,\eta}-e^{5/2\,\etap}))\frac{9}{2r}(r^2-1)^2\log\left|\frac{1+r}{1-r}\right|\Bigg]\,.
\label{SIGMA1L}
\eeqra
The general expression for the 1-loop contribution to $\Phi_{ab}^{(1)}(k;\eta,\etap)$ is
\beqra
&&\Phi_{ab}^{(1)}(k;\eta,\etap)= \nonumber\\
&&\qquad2 e^{\eta+\etap} \int d^3 q \gamma_{acd}(\bk,-\bq,-\bp) u_c P^0(q) u_e u_dP^0(p)u_f \gamma_{bef}(-\bk,\bq,\bp) \,,\nonumber\\
&&
\eeqra
with $\bp=\bk-\bq$, which gives the following expressions for the individual components,
\beqra
&&\Phi_{11}^{(1)}(k;\eta,\etap)=\nonumber\\
&&\qquad e^{\eta+\etap} \frac{ \pi}{4 k} \int_0^\infty dq\int_{|k-q|}^{k+q} dp\frac{\left[k^2(p^2+q^2)-(p^2-q^2)^2\right]^2}{p^3q^3} P^0(q)P^0(p)\,,\nonumber\\
&&\Phi_{12}^{(1)}(k;\eta,\etap)=\nonumber\\
&&\quad e^{\eta+\etap} \frac{k \pi}{4} \int_0^\infty dq\int_{|k-q|}^{k+q} dp\frac{(k^2-p^2-q^2)\left[k^2(p^2+q^2)-(p^2-q^2)^2\right]}{p^3q^3} P^0(q)P^0(p)\,,\nonumber\\
&&\Phi_{21}^{(1)}(k;\eta,\etap)=\Phi_{12}^{(1)}(k;\eta,\etap)\,,\nonumber\\
&&\Phi_{22}^{(1)}(k;\eta,\etap)=e^{\eta+\etap} \frac{k^3 \pi}{4} \int_0^\infty dq\int_{|k-q|}^{k+q} dp\frac{(k^2-p^2-q^2)^2}{p^3q^3} P^0(q)P^0(p)\,.
\label{phicomp}
\eeqra

\section{}

In this appendix we will give the explicit formulae to be inserted in the evolution equation for the nonlinear PS, eq.~(\ref{TReik}). The first line contains the matrix $\Omega_{ab}$, defined in eq.~(\ref{bigomega}). At the second line, the two functions $H_1(k;\eta,-\infty)$ and $H_2(k;\eta,-\infty)$ appear, which are given explicitly by
\beqra
H_1(k;\eta,-\infty)&=& \int_{-\infty}^\eta ds\, \Sigma_{1b}^{(1)}(k;\eta,s) u_b=\nonumber\\
&-&e^{2\eta}\,\frac{k^3\pi}{21}\int dr \Bigg[19-24r^2+9r^4-\nonumber\\
&& \qquad\qquad\frac{9}{2 r}(r^2-1)^3\log\left|\frac{1+r}{1-r}\right|\Bigg] P^0(kr)\,,\label{HH1}\\
H_2(k;\eta,-\infty)&=& \int_{-\infty}^\eta ds\, \Sigma_{2b}^{(1)}(k;\eta,s) u_b=\nonumber\\
&-&e^{2\eta}\,\frac{k^3\pi}{21}\int dr \Bigg[-\frac{9}{r^2}+52+9r^2-\nonumber\\
&& \qquad\qquad\frac{9}{2 r^3}(r^2-1)^3\log\left|\frac{1+r}{1-r}\right|\Bigg] P^0(kr)\,\label{HH2}.
\eeqra
As we have discussed in the text, the third line of eq.~(\ref{TReik}) is well approximated by using the small scale expression for $\tilde \Phi_{ab}$ given in eq.~(\ref{largephitnew}), and the small scale analytic expression for the propagator, $G_{ab}^{L}$, instead of the solution of eq.~(\ref{dgtot}) with eq.~(\ref{TRGL}), namely, $\bar G_{ab}$. The time integration can then be done analytically, to get.
\beqra
&& \int\, ds\;\big[\tilde{\Phi}_{ad}(k;\eta,s)  G_{bd}^{L}(k;\,\eta,s)+\bar G_{ad}^{L}(k;\,\eta,s)\tilde\Phi_{db}(k;s,\eta) \big] \nonumber\\
&& \qquad\qquad\qquad\qquad\qquad  = \tilde\Phi G^{A}_{ab}(k;\eta)+\tilde\Phi G^{B}_{ab}(k;\eta)\,,
\label{terza}
\eeqra
where
\beqra
 &&\tilde\Phi G^{A}_{11}(k;\eta)= \frac{\sqrt{\pi }\, \mathrm{Erf}(y)}{y}
   \left(\frac{3 \Phi^{(1)}_{11}(k;\eta,\eta)}{5}+\frac{2 \Phi^{(1)}_{12}(k;\eta,\eta)}{5}\right)\nonumber\\   
   &&\qquad\qquad+\frac{8}{525} \,{\cal B}(y^2)\,(\Phi^{(1)}_{11}(k;\eta,\eta)-\Phi^{(1)}_{12}(k;\eta,\eta))\,,\nonumber\\
   &&\tilde\Phi G^{A}_{12}(k;\eta)=\frac{\sqrt{\pi }\, \mathrm{Erf}(y)}{y}
   \left(\frac{3 \Phi^{(1)}_{11}(k;\eta,\eta)}{10}+\frac{\Phi^{(1)}_{12}(k;\eta,\eta)}{2}+\frac{\Phi^{(1)}_{22}(k;\eta,\eta)}{5}\right)\nonumber\\
 &&\qquad\qquad -\frac{2}{525} \, {\cal B}(y^2)\, (3\Phi^{(1)}_{11}(k;\eta,\eta) -5\Phi^{(1)}_{12}(k;\eta,\eta)+2\Phi^{(1)}_{22}(k;\eta,\eta))\,,\nonumber\\
 &&   \tilde\Phi G^{A}_{21}(k;\eta)= \tilde\Phi G^{A}_{12}(k;\eta)\,,\nonumber\\
&& \tilde\Phi G^{A}_{22}(k;\eta)=\frac{\sqrt{\pi } \,\mathrm{Erf}(y)}{y}
   \left(\frac{3 \Phi^{(1)}_{12}(k;\eta,\eta)}{5}+\frac{2 \Phi^{(1)}_{22}(k;\eta,\eta)}{5}\right)\nonumber\\
&&\qquad\qquad-\frac{4}{175} \, {\cal B}(y^2)\, (\Phi^{(1)}_{12}(k;\eta,\eta)-\Phi^{(1)}_{22}(k;\eta,\eta))\,,
\eeqra
and
\beq
 \tilde\Phi G^{B}_{ab}(k;\eta) = u_au_b P^0(k) \left[y^2(e^{-y^2}-1)+\sqrt{\pi}\, y^3  \mathrm{Erf}(y) \right]\,,
\eeq
with $y\equiv e^\eta\,k\,\sigma_v$.
\begin{figure}
\centerline{\includegraphics[width = 15cm,keepaspectratio=true]{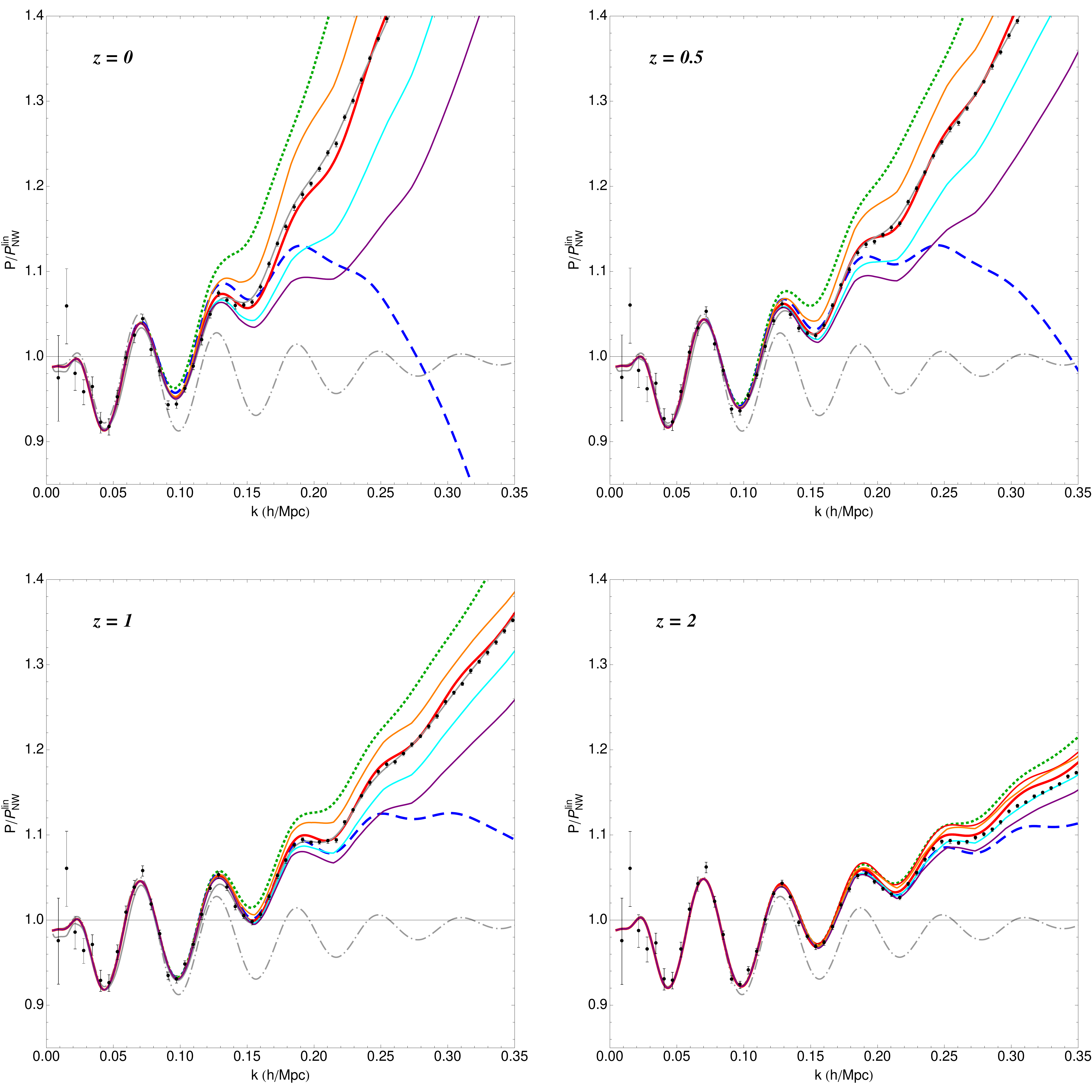}}
\caption{The results in the BAO region for $\bar k = 0.15,\; 0.20,\; 0.25,\;0.30\;\mathrm{h\,Mpc}^{-1}$, (orange, red, cyan, purple lines, respectively)}
\label{filt1}
\end{figure}
\label{AB}
\begin{figure}
\centerline{\includegraphics[width = 15cm,keepaspectratio=true]{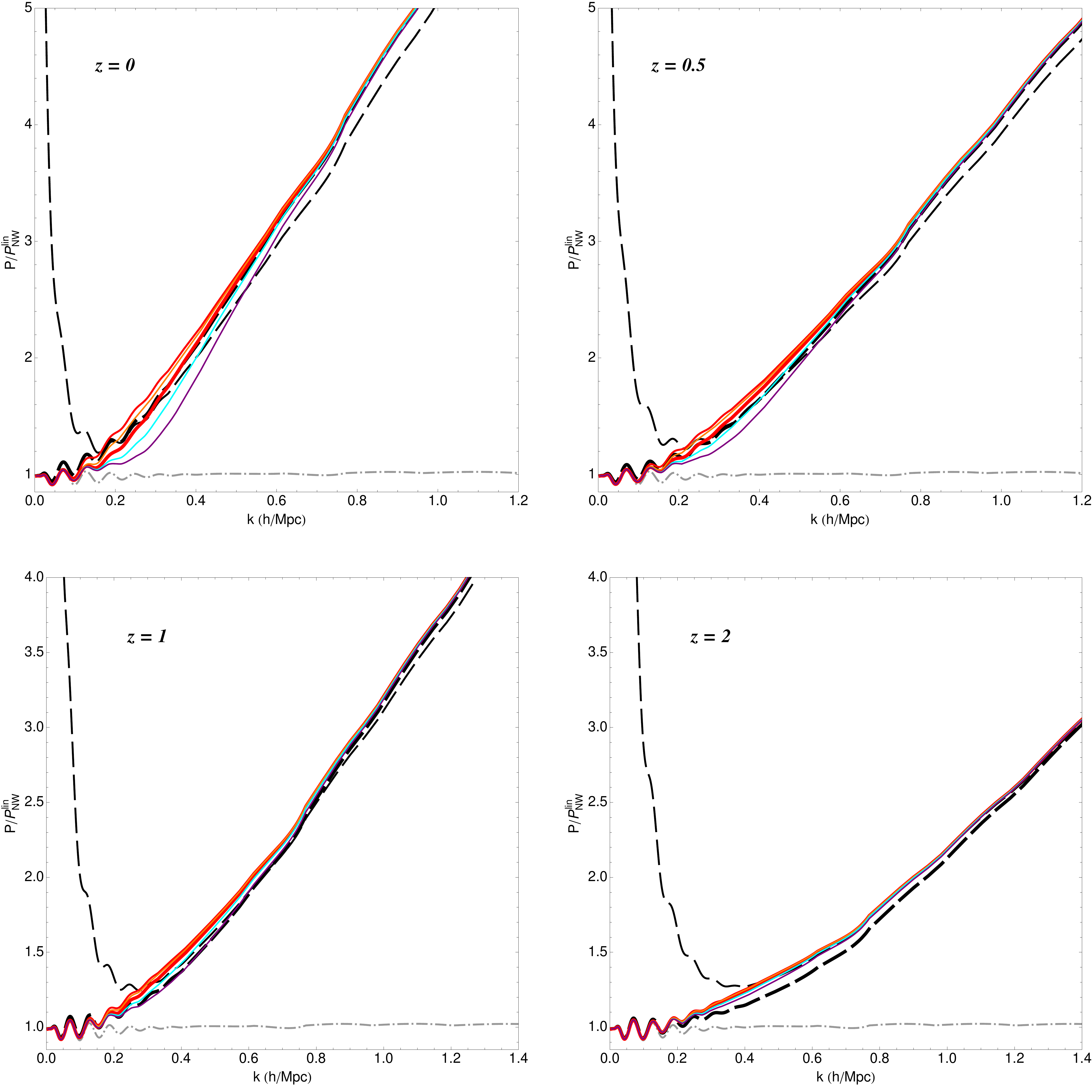}}
\caption{The results in the large $k$ region for $\bar k = 0.15,\; 0.20,\; 0.25,\;0.30\;\mathrm{h\,Mpc}^{-1}$, (orange, red, cyan, purple lines, respectively)}
\label{filt2}
\end{figure}
The function ${\cal B}(y^2)$ is a combination of generalized hypergeometric functions,
\beqra
&&{\cal B}(y^2) \equiv 35 \, _2F_2\left(\frac{1}{2},1;\frac{5}{4},\frac{7}{4};-y^2\right) -28  \, _2F_2\left(1,\frac{3}{2};\frac{7}{4},\frac{9}{4};-y^2\right)\nonumber\\
&&\qquad\qquad+ 8\,_2F_2\left(\frac{3}{2},2;\frac{9}{4},\frac{11}{4};-y^2\right)\,.
\eeqra
Eq.~(\ref{terza}) gives the large $y$ limit of the third line of eq.~(\ref{TReik}). In the very large $k$ limit $\Phi^{(1)}_{ab}(k;\eta,\eta)\to  u_a u_b P^0(k) y^2$ and eq.~(\ref{terza}) goes to eq.~\re{phig}.

As we have discussed in the text, in the small momentum limit the second term, $\tilde\Phi G^{B}_{ab}(k;\eta)$ has to be switched off, because it contains 2-loop expressions valid at large $k$. Therefore we will multiply it by a momentum cutoff function. In our numerical implementations, as discussed in the text, we have used the expression 
\beq
\ds
 \tilde\Phi G^{A}_{ab}(k;\eta)+ \frac{\left(\frac{k}{\bar k}\right)^4}{1+\left(\frac{k}{\bar k}\right)^4}\tilde\Phi G^{B}_{ab}(k;\eta)\,,
 \label{ufilter}
\eeq
where $\bar k = 0.2 \,\mathrm{h/Mpc}$ represents a reasonable value above which the large scale expression can start to be applicable. Alternatively, $\bar k$ can also be taken as a parameter to be marginalized in fits to real data. In figs.~\ref{filt1} and \ref{filt2} we show results obtained by changing the filter value from $\bar k =0.15$ to $0.30\;\mathrm{h\,Mpc}^{-1}$.

The initial conditions must be given at a very large redshift $z_{in}=O(100)$, where the PS can be approximated with the linear one:
\beq
P_{ab}(k;\etain) =u_a u_bP^0(k)\,, 
\eeq
and the equations can then be integrated town to the required final redshift, corresponding to $\eta_f = \log D(z_f)/D(z_{in})$.

\section*{References}
\bibliographystyle{JHEP}
\bibliography{/Users/pietroni/Bibliografia/mybib.bib}

\end{document}